\documentclass[twocolumn,showpacs,preprintnumbers,amsmath,amssymb]{revtex4}
\usepackage{epsfig}
\usepackage{amsthm}        
\usepackage{dcolumn}        
\usepackage{bm}
\usepackage{graphicx}
\usepackage{graphics}
\usepackage{subfigure}

\newcommand{\One}{\openone}

\begin{document}


\title{Model of response spectrum and modal interaction in coupled nanomechanical resonators
}

\author{J. Dorignac${}^1$, A. Gaidarzhy${}^1$, P. Mohanty${}^2$}

\affiliation{${}^1$ College  of  Engineering,  Boston University,  Boston  MA
02215\\
${}^2$ Department of Physics,  Boston University,  Boston  MA
02215}

\date{\today}
\begin{abstract}
We develop a simple {\em continuum} model to analyze the vibrational modes
of a nanomechanical multi-element structure. In this model, arrays
of sub-micron cantilevers located symmetrically on both sides of 
the central clamped-clamped nanobeam are replaced by a continuum. In this
approach, 
the punctual
shear forces exerted by the cantilevers on the central beam 
are smoothed out and the equations of motion of the structure become 
exactly solvable. Our analytical results capture the main
features of the vibrational modes observed both numerically and 
experimentally. Furthermore, 
using a perturbative approach to treat the nonlinear 
dynamics of the structure, we establish its frequency-amplitude response
and analyze the mechanism of anharmonic coupling between two specific
widely spaced modes of the resonator.   
\end{abstract}

\pacs{03.65.Ta, 62.25.-g, 62.30.+d, 62.40.+i}

\maketitle

\section{Introduction} \label{Introduction}

Nanomechanical resonators have been used  
to  investigate  fundamental  physics  problems  in a  wide  range  of
research  areas.   These   include  quantum  measurement  and  quantum
computation  \cite{wheeler,  zurek,  zurek07,  caves,  bocko-onofrio},
ultra-sensitive  force and  mass detection  \cite{rugar-force}, single
spin   detection  \cite{rugar-spin},   gravitational   wave  detection
\cite{caves, bocko-onofrio} and  other fundamental phenomena \cite{wu,
montemagno, braginsky, dorignac}. 
The central reason for the increasing interest
and  activity  in  nanomechanical   systems  for  quantum  studies  is
straightforward:  fast dynamics of  nanomechanical systems  enable the
investigation  of the  yet-unexplored  corner of  the parameter  space
where  ultra-high  frequency   resonance  modes  cooled  to  near-zero
temperatures are  expected to manifest  quantum mechanical corrections
to the classical motion \cite{gaidarzhy-prl, lahaye, Rajquantnanomech}.

Likewise, from a technological standpoint, compact size, robustness,
and  high frequencies  of nanomechanical  resonators have  resulted in
numerous  proposals  for  applications  where  technologies  involving
nanomechanical systems can offer alternative solutions to the existing
electronic   circuits  and   off-chip  devices:   frequency  selective
oscillators  \cite{nguyen1},  passive  filters \cite{nguyen2},  memory
elements     \cite{badzey-memory}      and     spintronics     devices
\cite{mohanty-prb}.

Despite  the  broad areas  of interest  in  nanomechanical systems,
comprehensive  studies of elasticity  and mechanics  of nanomechanical
structures  are  yet  to  be  done beyond  the  simple  single-element
structures such as cantilevers  and clamped beams. The main difficulty
has  been   the  extension  of  elasticity   theory  to  multi-element
structures for  closed-form analytical solutions. As  described in the
following, most of the eventual applications of nanomechanical systems
will   include  complex   structure   design.  Therefore,   analytical
understanding of  resonant modes and  other dynamical aspects  such as
nonlinear mode coupling in complex multi-element structures will be of
tremendous  importance.  Here, we  briefly  describe  a  few areas  of
interest  in which  nanomechanical structures  in the  gigahertz range
will play a key role.

\subsection{Entry into the Quantum Regime}

Experimental access to  the quantum realm is {\it  crudely} defined as
the regime in which the quantum of energy $hf$ in a resonant mode with
frequency  $f$  is  larger  than  the thermal  energy  $k_BT$  of  the
environment.  The  motivation  behind  this crude  definition  of  the
quantum regime is simple. The motion of a damped mechanical system can
be  described  by a  harmonic  oscillator  potential.  In the  quantum
regime,   the  harmonic   oscillator  potential   energy   levels  are
discrete.  In order  to  observe effects  of  discrete energy  levels,
smearing  by  thermal  energy---due   to  finite  temperature  of  the
mechanical  system---must  be  small  compared  to  the  energy  level
spacing, $hf$. The
condition  $hf  \ge  k_BT$  gives physically  relevant  parameters:  a
nanomechanical structure  with a normal mode resonance  frequency at 1
GHz  will enter  the  quantum  regime below  a  temperature $T  \equiv
(h/k_B)  f = 48~\mbox{mK}$.  Since typical  dilution cryostats  have a
base temperature of 10  mK, nanomechanical structures with frequencies
above 1 GHz can enable  experimental access to the quantum regime. The
experimental challenge is then to fabricate structures capable of high
gigahertz-range  resonance frequencies,  and  to measure  them at  low
millikelvin-range   temperatures.   Since   the  resonance   frequency
increases  with decreasing  size of  the system,  one or  many  of the
critical dimensions of the  gigahertz-frequency oscillators must be in
the sub-micron or nano scale.

\subsection{Mesoscopic Elasticity: Multiscale Modeling}

For  large mechanical  structures, continuum  mechanics  of elasticity
theory  provides  the appropriate  system  response,  both static  and
dynamic. As system  size shrinks down to the  submicron or nano scale,
the  elastic   behavior  starts   to  become  atomistic   rather  than
continuous,  and  it gives  rise  to  a  host of  anomalous  behaviors
\cite{maranganti}. These include enhancement of influence of 
surface defects, novel
dissipation  mechanisms,  reduction  of  the stiffness  constant,  and
statistical  fluctuation effects. Nanomechanical  resonators therefore
need to  be modeled  by atomistic simulation  in order to  capture the
essential  aspects   of  their  elastic  properties   arising  due  to
mesoscopic  size. However,  atomistic simulation  of  these structures
containing  about 100  million atoms  or more  becomes computationally
intensive.  Further problems arise  due to  the existence  of multiple
length  scales.  Modeling  of   structures  too  small  for  continuum
mechanics and  too large  for atomistic molecular  dynamics simulation
requires a proper understanding of the coupling of length scales.

For a  comprehensive study  of nanomechanical systems,  the structures
need    to    be   properly    characterized    for   their    elastic
behavior.  Specifically,  correlation   of  simulated  modeshapes  and
experimentally-measured modes  will require the  knowledge of relevant
static  and  dynamic  parameters.  Therefore,  analytical  studies  of
complex  nanomechanical  systems  will   not  only  provide  a  strong
motivation for  new approaches to  model materials beyond  the current
limit of computing  capacity, it will augment the  ongoing work on the
multiscale modeling of fracture dynamics and nanomaterials.

\subsection{Nanomechanical Device Applications in the Gigahertz Range} 

The driving  force behind MEMS  (Micro-Electro-Mechanical Systems), of
which NEMS  (Nano-Electro-Mechanical Systems) or  nanomechanics is the
natural  extension, has  been the  portfolio of  MEMS  applications in
optical communication (routers,  switches, repeaters), passive devices
in  cell-phone  industry   (filters,  accelerometers,  capacitors  and
inductors  for integrated  chip  design), and  sensor technologies  in
chemical,   biomedical,  and   electrical   solutions.  Nanomechanical
structures  are  faster  (because  of higher  resonance  frequencies),
therefore the  natural area of growth for  these applications includes
technologies     for     multi-function    nanomechanical     devices:
ultra-sensitive force sensors for  the detection of fundamental forces
and biomolecular forces. The technologies needed for observing quantum
effects  will have  obvious  use in  these  and other  gigahertz-range
applications.  In fact,  the range  of frequencies  discussed  in this
paper matches with  the frequency bands for communication  in a number
of consumer devices:  Cellular (0.4 GHz, 0.85 GHz,  1.9 GHz), WiFi and
Bluetooth  (2.4  GHz),  Satellite  radio  ($\sim$ 2.3  GHz),  and  GPS
receivers (L1-L3: 1.2-1.6 GHz).

In  basic research,  hybrid  nanomechanical devices  in the  gigahertz
range  can  prove to  be  crucial in  a  number  emerging fields.  For
example,  there  has been  a  recent  proposal  for spintronics  based
entirely on  nanomechanical torque  associated with the  electron spin
\cite{mohanty-prb}.  This  device  is  an  example  of  multi-function
nanomechanical  device, which,  in  the gigahertz  range, can  provide
access  to control  and  manipulation  of electron  spin  at the  spin
coherence and spin relaxation time scales of nanoseconds.

Since  the structure  size  is reduced  for  increasing resonant  mode
frequencies---and  hence the  operation  speed of  the  device, it  is
possible   that  within   a  few   years  certain   applications  with
nanomechanical devices  will emerge  which will have  frequencies high
enough   so  that   at   the  relevant   operating  temperatures   the
nanomechanical  element   will  be  quantum   mechanical.  A  specific
application  of  gigahertz-range  nanomechanical oscillators  involves
space  communication   devices  (0.5-18  GHz).   As  passive  devices,
nanomechanical structures in  certain space communication applications
are expected to  remain at low temperatures, perhaps  near the quantum
regime. This  leads to the obvious realization  that further shrinkage
of   micro-  or   nanomechanical  systems   with   corresponding  high
gigahertz-range frequencies will inevitably require new paradigms.

\subsection{Approaches to Gigahertz Resonance Frequencies}

Central to many of  the aforementioned applications of NEMS resonators
is the attainment high natural  frequencies of motion up to and beyond
the 1-GHz mark.  Among the various  solutions proposed  for achieving
ultra-high frequencies in nanomechanical resonators such as the use of
high   stiffness  materials  \cite{huang,   imboden}  and   bulk  mode
geometries  \cite{nguyen-bulk}, our  approach  of coupling  mechanical
elements  to  enhance  high  order  resonant modes  of  the  resonator
structure  has  been  shown  to  offer  a  number  of  advantages  for
performance and detection  ease \cite{gaidarzhy-apl}.  The coupling of
additional  degrees of  freedom  to a  10-$\mu$m-long simple  nanobeam
structure  can  produce significant  modification  of  the high  order
resonance  spectrum, with  enhanced amplitude  and quality  factors of
selected  vibration modes extending  well beyond  the 1  GHz frequency
range.

\subsection{Organization of the Paper}

Here,  we  present  two   analytical  models  that  yield  closed-form
solutions  describing  the  dynamics  of the  coupled-beam  resonator,
dubbed  the antenna  structure.  This  structure is  a prototype  of a
class  of two-element structures  that can  be envisioned  in specific
applications. The continuum model allows for a clear comparison of the
modal shapes  and spectrum  with full finite  element analysis  of the
structure.   We further investigate  nonlinear modal  coupling between
widely  spaced  modes  of  the  structure  using  perturbation  theory
techniques.

In  Section II,  we  describe  a discrete  model  of the  antenna-like
multi-element  structure,  and  we   obtain  exact  solution  for  the
vibrational  modes. In  Section  III,  we extend  this  analysis to  a
continuum  approximation   where  the  periodic   and  punctual  force
densities are smoothed  out. In this continuum model,  we describe the
frequency spectrum and the corresponding band structure. Specifically,
we  calculate fundamental  and  collective modes  and  we compare  the
results with  finite element  analysis. In Section  IV, we  consider a
specific  example of  nonlinear dynamics  of the  system in  which two
resonant  modes  are  coupled.  This  type of  mode  coupling,  as  we
describe in  detail, can allow detection  of response in  one mode by
monitoring the coupled mode.

\section{Discrete Model}
\subsection{Equations of motion} \label{SecIIsub1}

\begin{figure} 
\includegraphics[scale=0.7]{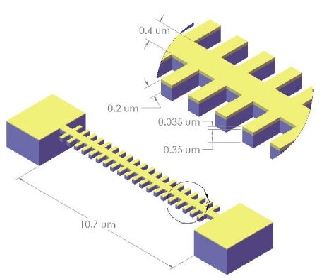} 
\caption{Antenna structure.} \label{Antenna}
\end{figure}
The antenna-like structure we investigate consists
of a central clamped-clamped diamond beam with dimensions $L \times W \times h = 10.7 \mu {\rm m} \times 0.4 \mu {\rm m} \times 0.35 \mu {\rm m}$ and 40 
perpendicular cantilevers with dimensions $l \times w \times h = 0.5 \mu {\rm m} \times 0.2 \mu {\rm m} \times 0.35 \mu {\rm m}$. 
The cantilevers are regularly spaced along the beam and symmetric with respect
to the beam (see Fig. \ref{Antenna}). There are $N=20$ cantilevers 
on each side. 
In addition, the antenna is coated by a layer of gold whose length 
and width are those of the beam and whose thickness is $t=0.035 \mu {\rm m}$.
We denote by $x$ the 
coordinate along the beam and by $\xi$ the transverse one. The deflection 
of the beam with respect to its clamps is called $y(x,t)$ and the
deflection of the $j$th lateral cantilever with respect to $y(x_j,t)$
is denoted by $\eta_j(\xi,t)$.
In the following we shall only consider vibrational modes symmetric with 
respect to the central beam. The deflection of the cantilevers
on either side of the beam are the same. They are described by the 
quantity $\eta_j(\xi,t)$. Considering that all the elements of the structure 
are one-dimensional and using the Euler-Bernoulli beam theory 
(see for example Ref. \cite{Segel,Gere,Nayfeh04}), 
the equations of motion of the structure have the form
\begin{eqnarray}
{\cal E}_b \frac{\partial^4 y}{\partial x^4} + \mu_b 
\frac{\partial^2 y}{\partial t^2} = - 2 {\cal E}_t 
\sum_{j=1}^{N} \left. \frac{\partial^3 \eta_j}{\partial \xi^3} 
\right|_{\xi=0} \delta(x-x_j) \label{y1}\\
{\cal E}_t \frac{\partial^4 \eta_j}{\partial \xi^4} + \mu_t 
\frac{\partial^2 \eta_j}{\partial t^2} = -\mu_t 
\frac{\partial^2 y_j}{\partial t^2},\ \ j \in \{1,\dots,N\} \label{eta1} 
\end{eqnarray}  
where $y_j\equiv y(x_j,t)$ and together with the boundary conditions 
\begin{eqnarray}
y(0,t)=\frac{\partial y}{\partial x}(0,t)=y(L,t)=\frac{\partial y}{\partial x}(L,t)=0 \label{bcy1}\\
\eta_j(0,t)=\frac{\partial \eta_j}{\partial \xi}(0,t)=
\frac{\partial^2 \eta_j}{\partial \xi^2}(l,t)=
\frac{\partial^3 \eta_j}{\partial \xi^3}(l,t)=0, \label{bceta1}
\end{eqnarray}
with $j \in \{1,\dots,N\}$.    
The rigidities of the beam, ${\cal E}_b$, and the cantilevers, ${\cal E}_t$ - 
that account for the presence of the gold layer - are respectively given by 
(see e.g. Ref. \cite{Gere})
\begin{equation} 
{\cal E}_b = \frac{W}{12}\frac{\left[ E_d^2 h^4 + E_d E_g 
(4 h^3 t + 6 h^2 t^2 + 4 h t^3) + E_g^2 t^4\right]}{(E_d h +E_g t)}\nonumber 
\end{equation}
\begin{equation}\label{EtvsEb}
{\cal E}_t = \frac{w}{W} {\cal E}_b.
\end{equation}
where $E_d=700\, {\rm GPa}$ and $E_g=45.6\, {\rm GPa}$ are the measured 
Young's moduli of diamond and gold, respectively.
The masses per unit length of the beam and a cantilever are respectively 
given by
\begin{equation} \label{mutvsmub}
\mu_b = W (\rho_d h + \rho_g t) \ \ ;\ \ \mu_t = \frac{w}{W} \mu_b 
\end{equation}
where $\rho_d=4050\, {\rm kg.m}^{-3}$ and $\rho_g=19500\, {\rm kg.m}^{-3}$ 
are the densities of diamond and gold, respectively. It is important to 
note that 
because of the smallness of $E_g$ with respect to $E_d$, the rigidities
of the beam and the cantilever are barely affected by the presence of the 
gold layer 
(roughly 2.5 \% higher). This is in contrast with their linear mass 
that the gold
layer increases by roughly 50 \%. Thus, the presence of the gold layer 
merely increases the mass of the antenna without affecting its rigidity.

The force density in the r.h.s of equation \eqref{y1} represents the shear
force densities exerted by the cantilevers on the beam. No momenta appear in
this equation because, for modes that are symmetric with
respect to the central beam, the momenta exerted by two opposite 
cantilevers cancel each other. The r.h.s.
of equation \eqref{eta1} is due to the motion of the base of the 
cantilever that 
follows the motion of the central beam at $\xi=0$.

\subsection{Energy and Lagrangian}

The system of equations \eqref{y1} and \eqref{eta1} (for $j \in \{1,\dots,N\}$) conserves the energy 
\begin{multline} \label{H1}
H_d = \int\limits_0^L dx \left[\frac{{\cal E}_b}{2}\left(\frac{\partial^2 y}{\partial x^2}\right)^2 +
\frac{\mu_b}{2}\left(\frac{\partial y}{\partial t}\right)^2 
\right] + \\ 
2 \sum_{j=1}^N \int\limits_0^l d\xi 
\left[\frac{{\cal E}_t}{2}\left(\frac{\partial^2 \eta_j}{\partial \xi^2}
\right)^2 
+\frac{\mu_t}{2}\left(\frac{\partial \eta_j}{\partial t}+
\frac{\partial y_j}{\partial t}\right)^2 \right].
\end{multline} 
The first and second terms in Eq. \eqref{H1} represent 
the elastic and kinetic energies of the central beam and of the 
$2N$ cantilever,
respectively. Similar to Hamiltonian \eqref{H1}, the Lagrangian of 
the antenna, ${\cal L}_d$, is expressed as 
\begin{multline} \label{L1}
{\cal L}_d = \int\limits_0^L dx \left[\frac{\mu_b}{2}\left(\frac{\partial y}{\partial t}\right)^2 -\frac{{\cal E}_b}{2}\left(\frac{\partial^2 y}{\partial x^2}\right)^2 
\right] + \\ 2 \sum_{j=1}^N \int\limits_0^l d\xi 
\left[\frac{\mu_t}{2}\left(\frac{\partial \eta_j}{\partial t}+
\frac{\partial y_j}{\partial t}\right)^2 -\frac{{\cal E}_t}{2}\left(\frac{\partial^2 \eta_j}{\partial \xi^2}
\right)^2 \right].
\end{multline}
Eqs. \eqref{y1}-\eqref{eta1} can be 
established from the least action
principle $\delta {\cal S}_d = 0$, i.e. $\frac{\delta {\cal S}_d}{\delta y}=0$ 
and $\frac{\delta {\cal S}_d}{\delta \eta_j}=0$, where the action is given by 
${\cal S}_d = \int \! {\cal L}_d \, dt$.  

\subsection{Vibrational modes} \label{SecDiscVibMod}

The vibrational modes of Eqs. \eqref{y1}-\eqref{bceta1} can 
be found as follows. 
Given the punctual nature of the force densities,
Eq. \eqref{y1} reduces to 
\begin{equation}
{\cal E}_b \frac{\partial^4 y}{\partial x^4} + \mu_b 
\frac{\partial^2 y}{\partial t^2} = 0\ \ \text{for}\ \ 
x \in [0,L], x \neq x_j, \label{y1b}
\end{equation}
with $j \in \{1,\dots,N\}$.
At all points $x_j$, 
the function $y(x,t)$ and its first and second derivatives are continuous.
But its third derivative is discontinuous and its discontinuity 
is obtained by integrating Eq. (1) over an infinitesimally small interval 
centered around $x_j$. Thus, in addition to the boundary 
conditions given in Eq. \eqref{bcy1}, $4N$ other conditions apply:
the continuity of $y(x,t)$ and its first and second derivatives at 
$x=x_j$, $j \in \{1,\dots,N\}$, and
\begin{equation}
{\cal E}_b \left(\frac{\partial^3 y}{\partial x^3}(x_j^+,t)-
\frac{\partial^3 y}{\partial x^3}(x_j^-,t)\right) = -2{\cal E}_t
\frac{\partial^3 \eta_j}{\partial \xi^3}(0,t).
\label{bcy1b}
\end{equation}
To simplify the system of Eqs. \eqref{y1}-\eqref{eta1}, as well as the 
boundary conditions, we shall work with 
the following non-dimensional quantities 
\begin{multline} \label{nondim}
u=\frac{x}{L}\ ;\ v=\frac{\xi}{l}\ ;\ 
\mu^4 = \frac{\mu_b L^4 \omega^2}{{\cal E}_b}\ ;\\ 
\gamma^4 = \frac{\mu_t l^4 \omega^2}{{\cal E}_t}\ ;\
R = 2 \left(\frac{L}{l}\right)^3 \frac{{\cal E}_t}{{\cal E}_b}. 
\end{multline} 
Looking for mode solutions on the form $y(x,t)=Y(u)\cos(\omega t)$ and 
$\eta_j(\xi,t)=H_j(v)\cos(\omega t)$, the equations to be solved become
\begin{eqnarray}
&& \frac{d^4 Y(u)}{d u^4} -\mu^4 Y(u) = 0\ \ \text{for}\ \ 
u \in [0,1], u \neq u_j, \label{Yudiseq}\\
&& \frac{d^4 H_j(v)}{dv^4} -\gamma^4 
(H_j(v)+Y(u_j))=0, \label{Hjvdiseq}
\end{eqnarray}
and the boundary conditions \eqref{bcy1}, \eqref{bceta1} and \eqref{bcy1b}
read
\begin{eqnarray} \label{BCcant}
&& \hskip-5ex Y(0)=Y'(0)=Y(1)=Y'(1)=0, \nonumber \\ 
&& \hskip-5ex Y(u_j^-)=Y(u_j^+)\ ;\ 
Y'(u_j^-)=
Y'(u_j^+) ; \nonumber \\ 
&& \hskip-5ex Y''(u_j^-)=
Y''(u_j^+) \ ; \
Y'''(u_j^+)-Y'''(u_j^-) = -R
H_j'''(0), \nonumber \\ 
&&  
\hskip-5ex H_j(0)=H_j'(0)=H_j''(1)=H_j'''(1)=0, 
\end{eqnarray} 
where $u_j = j/(N+1)$, $j \in \{1,\dots,N\}$.
The solution to the system of equations \eqref{Yudiseq}-\eqref{Hjvdiseq} is
worked out in appendix \ref{Gensoldisc}. From it, we eventually obtain 
the secular equation of the discrete model 
\begin{equation} \label{SecEqDisc} 
\det M(\omega) = 0,
\end{equation}
where ${\bf M}(\omega)$ is a $2 \times 2$ matrix given by
\begin{equation} \label{MatM}
{\bf M}(\omega) = {\bf T}(\mu) \left[ \prod_{j=1}^N (\One + 
2\alpha A_2(\gamma) {\bf K_j}(\mu)) \right] {\bf L}.
\end{equation} 
In the expression above, $\One$ is the $4 \times 4$ unity matrix, $\alpha=w/W$,
$A_2(\gamma)$ is defined in Eq. \eqref{Aju}. The quantities 
$\mu$ and $\gamma$, defined in \eqref{nondim}, are related to each 
other as $\mu=\gamma L/l$. They are the frequency dependent parameters
of Eq. \eqref{MatM}. Finally, the matrices ${\bf K_j}(\mu)$, ${\bf T}(\mu)$ 
and ${\bf L}$ are given by
\begin{equation} 
{\bf K_j}(\mu) =  
\begin{pmatrix}
s_j c_j & s_j^2 & s_j.ch_j & s_j.sh_j \\
-c_j^2 & -c_js_j & -c_j.ch_j & -c_j.sh_j \\
-c_j.sh_j & -s_j.sh_j & -sh_j.ch_j & -sh_j^2 \\
c_j.ch_j & s_j.ch_j & ch_j^2 & ch_j.sh_j \\
\end{pmatrix} \nonumber
\end{equation}
\begin{equation} \label{MatKjLT}
{\bf L} =  
\begin{pmatrix}
1 & 0 \\
0 & 1 \\
-1 & 0 \\
0 & -1 \\
\end{pmatrix}\ ;\ 
{\bf T}(\mu) =  
\begin{pmatrix}
\cos \mu & -\sin \mu \\
\sin \mu & \cos \mu \\
\cosh \mu & \sinh \mu \\
\sinh \mu & \cosh \mu \\
\end{pmatrix}^T. 
\end{equation} 
where ${\bf A}^T$ denotes the transpose of ${\bf A}$ and where 
$ c_j\equiv \cos(\mu u_j)\ ;\ s_j\equiv \sin(\mu u_j)\ ;\  
ch_j\equiv \cosh(\mu u_j)\ ;\  sh_j\equiv \sinh(\mu u_j)\ ;\ 
u_j=j/(N+1).$ 
Using this, we obtain the exact solution
of the discrete model. Nonetheless, the method is particularly 
tedious in practice
and it 
does not allow for an easy analytical investigation of the problem unless
the number of cantilevers to be treated is very small. The case $N=1$ is 
treated in appendix \ref{App2tdeltavscont} and compared to the result
given by the continuum approach developed hereafter. 

\section{Continuum approximation}

\subsection{Derivation of the model}

We now derive a continuum approximation for the system of Eqs. 
\eqref{y1}-\eqref{bceta1}. The idea is to ``smooth out'' the punctual
force densities of Eq. \eqref{y1} in such a way that the total shear 
force exerted by the $j$th cantilever on the beam is the same but 
with a density
that is continuous along the beam. The shear force exerted by the two 
cantilevers in the $j$th position is given by 
$F_j = -2{\cal E}_t (\partial^3 \eta_j/\partial \xi^3)(0,t)$. Averaging 
its effect over the interval $[x_j-\delta/2,x_j+\delta/2]$ produces the force 
density $f_j=F_j/\delta$ and filling the gap between the cantilevers 
requires $\delta=L/N$. Using this, we obtain a piecewise-constant
force density given by $f_j = -(2N{\cal E}_t/L) 
(\partial^3 \eta_j/\partial \xi^3)(0,t)$ whose continuous version 
is straightforward if we now assume that the cantilevers form a continuum
along the beam. Writing $\eta_j(\xi,t)\equiv \eta(x_j,\xi,t)$, the 
continuous version of the density simply reads $f(x)=-(2N{\cal E}_t/L) 
(\partial^3 \eta/\partial \xi^3)(x,0,t)$. Therefore, the equations of motion
become
\begin{eqnarray}
&& {\cal E}_b \frac{\partial^4 y(x,t)}{\partial x^4} + \mu_b 
\frac{\partial^2 y(x,t)}{\partial t^2} = - \frac{2 {\cal E}_t N}{L}
\left. \frac{\partial^3 \eta(x,\xi,t)}{\partial \xi^3} 
\right|_{\xi=0}  \label{y2}\\
&& {\cal E}_t \frac{\partial^4 \eta(x,\xi,t)}{\partial \xi^4} + \mu_t 
\frac{\partial^2 \eta(x,\xi,t)}{\partial t^2} = -\mu_t 
\frac{\partial^2 y(x,t)}{\partial t^2}, \label{eta2} 
\end{eqnarray}  
with clamped-clamped and cantilever-like boundary conditions for the central 
beam and the cantilevers, respectively 
\begin{eqnarray}
&& \hskip-7ex y(0,t)=\frac{\partial y}{\partial x}(0,t)=y(L,t)=\frac{\partial y}{\partial x}(L,t)=0 \label{bcy2}\\
&& \hskip-7ex \eta(x,0,t)\!=\!\frac{\partial \eta}{\partial \xi}(x,0,t)\!=\! 
\frac{\partial^2 \eta}{\partial \xi^2}(x,l,t)\!=\!
\frac{\partial^3 \eta}{\partial \xi^3}(x,l,t)\!=\!0. \label{bceta2}
\end{eqnarray}  
This model is of course expected to provide better results as the density 
of cantilevers increases.
Equations \eqref{y2}-\eqref{eta2} conserve the following energy
\begin{multline} \label{H2}
H_c = \int\limits_0^L dx \left[\frac{{\cal E}_b}{2}\left(\frac{\partial^2 y}{\partial x^2}\right)^2 +\frac{\mu_b}{2}\left(\frac{\partial y}{\partial t}\right)^2 
\right] + \\ \frac{2N}{L} \int\limits_0^L dx \int\limits_0^l d\xi 
\left[\frac{{\cal E}_t}{2}\left(\frac{\partial^2 \eta}{\partial \xi^2}
\right)^2 
+\frac{\mu_t}{2}\left(\frac{\partial \eta}{\partial t}+
\frac{\partial y}{\partial t}\right)^2 \right].
\end{multline} 
Similar to the equations of the discrete model \eqref{y1}-\eqref{eta1}, Eqs.
\eqref{y2}-\eqref{eta2}
can be deduced from the least action principle $\delta {\cal S}_c = 0$, where 
${\cal S}_c = \int \! {\cal L}_c \, dt$ and where the Lagrangian ${\cal L}_c$
is obtained from Hamiltonian \eqref{H2} by changing the sign of the elastic
energy terms. 

The continuum model presented above becomes more 
accurate as the density of cantilevers, $N/L$, increases. To get a sense 
of its limitations though, let us imagine that, for a given mode, the shape
of the central beam has $n-1$ nodes, and consequently $n$ ``arches''. 
We then expect that the model holds if at least 
one cantilever per arch subsists, which basically yields the following 
condition of validity of the continuum model,
\begin{equation}\label{Validcond}
n < N.
\end{equation}
Indeed, should condition \eqref{Validcond} fail to be satisfied,  
the ``cantilever continuum'' would create some inertial effects 
on an arch where no physical cantilever is to be found, which is not
desirable. 
The limit of a single cantilever per arch, instead of two or more, 
can still appear rather arbitrary.
But a comparison of the lower part of the spectrum of the 
continuum and discrete models for $N=1$ done in appendix 
\ref{App2tdeltavscont} reveals that their first frequencies are 
very similar. Therefore, thinking of the central beam as a collection 
of sub-beams (arches) pinned at the level of its nodes, it seems reasonable 
to take one cantilever per arch as the limit of validity of the 
continuum model.

\subsection{Vibrational modes}

Working again with the non-dimensional quantities defined in Eq. 
\eqref{nondim},
we look for mode solutions of the form    
\begin{equation} \label{modeyeta}
y(x,t) = Y(u) \cos(\omega t)\ \ ;\ \  \eta(x,\xi,t)= H(u,v) \cos(\omega t).
\end{equation}
The equations of motion \eqref{y2} and \eqref{eta2} become
\begin{eqnarray}
&& \frac{d^4 Y(u)}{du^4} - \mu^4 Y(u) = - R N
\frac{\partial^3 H}{\partial v^3}(u,0)  \label{Yu2}\\
&& \frac{\partial^4 H(u,v)}{\partial v^4} - 
\gamma^4 \left(H(u,v)+Y(u)\right) = 0 \label{Huv2} 
\end{eqnarray}  
with boundary conditions 
\begin{eqnarray}
&& \hskip-7ex Y(0)=\frac{dY}{du}(0)=Y(1)=\frac{dY}{du}(1)=0 \label{bcYu2}\\
&& \hskip-7ex H(u,0)=\frac{\partial H}{\partial v}(u,0)=
\frac{\partial^2 H}{\partial v^2}(u,1)=
\frac{\partial^3 H}{\partial v^3}(u,1)=0. \label{bcHuv2}
\end{eqnarray}  
Now, solving Eq. \eqref{Huv2} yields
\begin{eqnarray} \label{Huvsol}
H(u,v) &=& \left[A_1(\gamma) \cos(\gamma v) + A_2(\gamma)\sin(\gamma v) + 
\right. \nonumber \\
&& \left.
A_3(\gamma)\cosh(\gamma v) + A_4(\gamma) \sinh(\gamma v) - 1\right]Y(u) 
\nonumber \\
&\equiv& \tilde{H}(v) Y(u),
\end{eqnarray}
where the coefficients $A_i(\gamma)$, determined by the boundary 
conditions \eqref{bcHuv2}, are given by
\begin{eqnarray}\label{Aju} 
A_1(\gamma) &=&  \frac{1+\cos \gamma \cosh \gamma - 
\sin \gamma \sinh \gamma}{2(1+\cos \gamma \cosh \gamma)}, \nonumber \\  
A_2(\gamma) &=& \frac{\cos \gamma \sinh \gamma + 
\sin \gamma \cosh \gamma}{2(1+\cos \gamma \cosh \gamma)},  \nonumber \\
A_3(\gamma) &=& \frac{1+\cos \gamma \cosh \gamma + 
\sin \gamma \sinh \gamma}{2(1+\cos \gamma \cosh \gamma)}, \nonumber \\
A_4(\gamma) &=& - \frac{\cos \gamma \sinh \gamma + 
\sin \gamma \cosh \gamma}{2(1+\cos \gamma \cosh \gamma)}. 
\end{eqnarray}
From these results we can calculate the r.h.s. of Eq. \eqref{Yu2}, $- RN 
\partial^3 H/\partial v^3(u,0)= 2 R N \gamma^3 A_2(\gamma)Y(u)$, and we 
finally obtain
\begin{equation} \label{Yueq}
\frac{d^4 Y(u)}{du^4} - \beta^4 Y(u) = 0, 
\end{equation}
with
\begin{equation} \label{beta4}
\beta^4 = \mu^4 + \gamma^3 R N \frac{\cos \gamma \sinh \gamma + 
\sin \gamma \cosh \gamma}{1+\cos \gamma \cosh \gamma}.
\end{equation}
Eqs. \eqref{Yueq} and \eqref{beta4} 
show that our continuum model reduces to a simple clamped-clamped beam
equation with a parameter $\beta$ that depends in a non trivial way upon
the mode frequency $\omega$. Indeed, according to Eq. \eqref{nondim}, the 
parameters $\mu^4$ and $\gamma^4$ are quadratic functions of the frequency
and the parameters $R$ and $N$ are simply constants. 
Now, because Eq. \eqref{Yueq} 
along with boundary conditions \eqref{bcYu2} is merely 
a clamped-clamped beam equation, it can be solved to obtain
\begin{multline} \label{Yusol}
Y(u) = A\Big\{\cos(\beta u) - \cosh(\beta u)  \\ 
-  \left(\frac{\cos \beta  
- \cosh \beta}{\sin \beta  - \sinh \beta }\right) 
(\sin(\beta u) - \sinh(\beta u))\Big\},
\end{multline}
with the secular equation
\begin{equation} \label{Eqbeta}
\cos(\beta ) \cosh(\beta) = 1.
\end{equation}

\subsection{Frequency spectrum of the continuum model: Band structure} 
\label{SecFreqSpec}

\begin{figure}
\includegraphics[height=.47\textwidth,width=.47\textwidth]{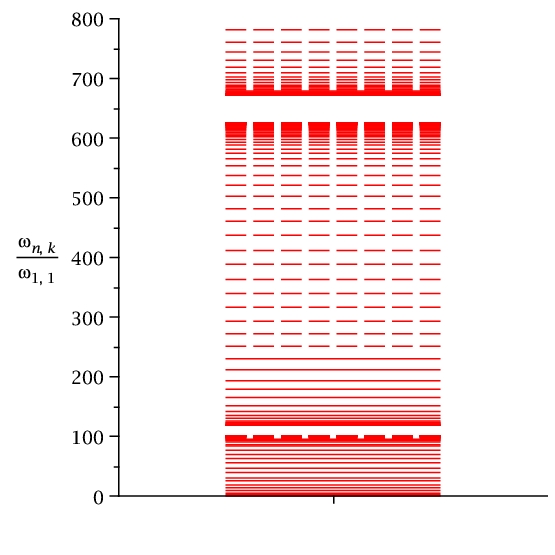}
\caption{Frequency spectrum obtained from Eq. \eqref{secEq} 
for $N=20$ and the antenna parameters given in \ref{SecIIsub1}. Solid lines 
correspond to $n<N$ (physically relevant frequencies) and dashed ones to 
$n>N$.}\label{Fig_Freq_spectrum}
\end{figure}

We see that
the shape of the vibrational modes
of the central beam is uniquely determined by the parameter $\beta$ solution
to Eq. \eqref{Eqbeta}. The solutions $\beta_n$ can be evaluated numerically:
$\beta_1=4.730040745$, $\beta_2=7.853204624$, $\beta_3=10.99560784$, 
and $\beta_n\simeq (n+1/2)\pi$ as $n$ is large. To a given $\beta_n$ 
corresponds a single modal shape for the central beam, $Y_n(u)$, 
but an infinite number of modal shapes $H_{n,k}(u,v)$ and frequencies 
$\omega_{n,k}$, $k\geq 1$, obtained
by solving Eq. \eqref{beta4} for $\omega$ . To see that the 
number of solutions 
to Eq. \eqref{beta4} for $\beta=\beta_n$ is infinite, we first rewrite
it in terms of the variable $\gamma$ by using 
$\mu= \gamma L/l$ and denoting by $m_b=L\mu_b$ and $m_t=l\mu_t$ the masses of the beam and of one cantilever, respectively, we finally find 
\begin{multline} \label{secEq}
Q(\gamma,\beta_n) \equiv \left(\cos \gamma+\frac{1}{\cosh \gamma}\right)
\left[1-\left(\frac{l\beta_n}{L\gamma}\right)^4 \right]\\ 
+ \frac{2N m_t}{m_b}\frac{\cos \gamma \tanh \gamma + \sin \gamma}{\gamma}= 0
\end{multline}
valid for $\cos \gamma \cosh \gamma +1 \neq 0$. It becomes now clear that 
when $\gamma$ is large, Eq. \eqref{secEq} reduces to $\cos \gamma=0$.
This equation has an infinite number of solutions with 
asymptotic behavior $\gamma_{n,k} = (k-1/2)\pi$ as $k \rightarrow \infty$. 
Interestingly, these asymptotic solutions are independent
of all the parameters entering Eq. \eqref{secEq}, and in particular, of 
$\beta_n$ {\em provided the latter is finite, i.e. $n \nrightarrow \infty$}. 
Finally, the frequency corresponding to a given value $\gamma_{n,k}$ 
is given by Eq. \eqref{nondim} as
\begin{equation} \label{Omega}
\omega_{n,k} = 
\sqrt{\frac{{\cal E}_t}{\mu_t}}\left(\frac{\gamma_{n,k}}{l}\right)^2.
\end{equation}
To understand the structure of the spectrum, we have plotted in Fig. 
\ref{Fig_Freq_spectrum} the frequencies $\omega_{n,k}$ 
normalized to the fundamental, $\omega_{1,1}$, 
for $N=20$ and the antenna dimensions indicated in \ref{SecIIsub1}.
\begin{figure}
\includegraphics[width=.47\textwidth]{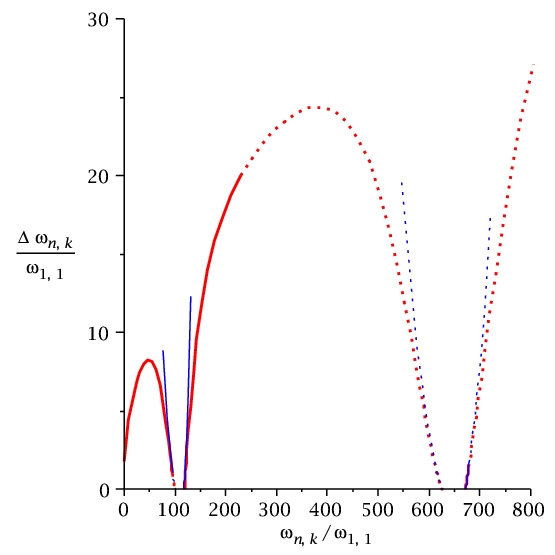}
\caption{Frequency spacing, 
$(\omega_{n+1,k}-\omega_{n,k})/\omega_{1,1}$, 
versus frequency (in red) for the same parameters as in Fig. 
\ref{Fig_Freq_spectrum}.  
In blue, analytical results obtained from Eq. \eqref{gammamidband}. Curves 
are solid for $n<N$ and dotted for $n>N$.}\label{Fig_Spac_vs_Freq_N20}
\end{figure}
As we can see, the spectrum consists of ``bands'' separated by gaps.
The band number is the label $k$ that we have attributed to the solution
of Eq. \eqref{secEq}, $\gamma_{n,k}$. To elucidate the appearance of bands,
we note that as $n$ increases, the 
solution to Eq. \eqref{Eqbeta} becomes large and is asymptotically 
given by $\beta_n \sim (n+1/2)\pi$. For a finite solution, $\gamma_{\infty,k}$,
of Eq. \eqref{beta4} to exist as $n \rightarrow \infty$, the denominator 
of the second term of the r.h.s of \eqref{beta4} needs to vanish. Hence,
\begin{equation} \label{seccanti} 
1+\cos(\gamma_{\infty,k})\cosh(\gamma_{\infty,k})=0.
\end{equation}
This equation is the well-known secular equation 
for a simple cantilever. It provides here the upper band 
edge of the $k$th band whose frequency, determined by Eq. \eqref{Omega},
reads
\begin{equation} \label{Omega_infty}
\omega_{\infty,k} = 
\sqrt{\frac{{\cal E}_t}{\mu_t}}\left(\frac{\gamma_{\infty,k}}{l}\right)^2.
\end{equation}
For the case at hand, we have $\omega_{\infty,1}/\omega_{1,1} \simeq 100.103$,
$\omega_{\infty,2}/\omega_{1,1} \simeq 627.336$ and 
$\omega_{\infty,k}/\omega_{1,1} \simeq ((k-1/2)\pi/\gamma_{1,1})^2$ with
$\gamma_{1,1}\simeq 0.187413$ for $k\geq 3$. It is easy to see that the 
accuracy of the latter formula improves exponentially as $k$ increases. For 
$k=3$ the relative error to the exact result is already as small as 0.02\%.
An approximate analytical expression for $\gamma_{1,1}$ is given in the next
section while results for the lower edge of the $k$th
band, $\gamma_{1,k}$, $k\geq 2$, will be given in section \ref{SecFreqClus}. 

At the edges of each band but the lower edge of the first one 
that we shall refer to as the ``fundamental band'', frequencies 
clearly accumulate. However,
close to the lower band edge, a finite number of frequencies cluster while
an infinite number accumulate at the upper band edge. Each band contains
all possible values of $\beta_n$, $1 \leq n \leq \infty$, i.e. all possible 
modal shapes for the central beam, $Y_n(u)$. Even though modes with 
the same shape $Y_n(u)$ repeatedly appear within each band, they differ
from band to band because their cantilever continuum, $H_{n,k}(u,v)$, 
depends on $\gamma_{n,k}$ that is, on
both $n$ and $k$ (see Eq. \eqref{Huvsol}). Note also that, 
within a given band, the frequency increases 
with the excitation level of the central beam. 

At mid-band, modal frequencies are somewhat sparse and  
more regularly spaced than at the boundaries. This is confirmed 
in Fig. \ref{Fig_Spac_vs_Freq_N20} that displays the normalized 
spacing between two consecutive 
frequencies versus the normalized frequency itself. Notice that, the inverse 
of this function is nothing but the normalized ``density of states'' of the 
spectrum. Mid-band frequencies can be approximately evaluated
once we note that the function $A_2(\gamma)$
appearing in secular equation \eqref{beta4} 
is small away from the boundaries $\gamma_{\infty,k}$ given by 
Eq. \eqref{seccanti} and increases slowly and regularly in this 
region. Rewriting Eq. \eqref{beta4} as 
$\mu^4-\beta_n^4+4\alpha N A_2(\mu l/L)\mu^3=0$, we seek a solution
close to $\mu \sim \beta_n$. With a first Newton iteration we obtain 
\begin{multline} \label{gammamidband}
\hskip-2ex \gamma_{n,k} \simeq \gamma_n^{(0)} \left[1 - 
\frac{\alpha N A_2(\gamma_n^{(0)})}
{\beta_n+3 \alpha N A_2(\gamma_n^{(0)})
+ \alpha N \beta_n \frac{l}{L} A'_2(\gamma_n^{(0)})}
\right],\\
\left| \beta_n - \mu_k^{*}\right| \ll \frac{L \Delta_k}{2l}.
\end{multline}
where $\gamma_n^{(0)} = \frac{\beta_n l}{L}$, 
$\alpha = w/W$ and $A_2(\gamma)$ is given in Eq. \eqref{Aju}.
In the last equation, $\Delta_k=\gamma_{\infty,k}-\gamma_{1,k}$, 
is the width of band $k$ and 
$\mu_k^{*}$ is determined as the solution to 
$A_2(\mu_k^{*}l/L)=0$. For $k$ large enough, 
$\mu_k^{*} \simeq (k-1/4)\pi L/l$, $k \geq 1$. In this case, we have also
$\beta_n\simeq (n+1/2)\pi$, $n \geq 1$. Approximating the bandwidth by 
$\Delta_k \simeq \gamma_{\infty,k}-\gamma_{\infty,k-1} \simeq \pi$ as $k$
is large enough, the condition of Eq. 
\eqref{gammamidband} becomes $|(n+1/2)l/L-(k-1/4)| \ll 1/2$, which 
provides bounds on the beam excitation $(n)$ of band $k$ for which
\eqref{gammamidband} is valid. In this case Eq. \eqref{gammamidband}
simplifies to 
\begin{equation} \label{gammamidbandasympt}
\gamma_{n,k} \simeq \frac{l}{L}\left[\beta_n - 
\alpha N A_2\left(\frac{\beta_n l}{L}\right)\right],\ \ \ 
\left| \beta_n - \mu_k^{*}\right| \ll \frac{L \Delta_k}{2l}.
\end{equation}
Although simple, this last expression 
is not very accurate for the first bands. 

Very few of the many frequencies occurring in the spectrum of the continuum
model can be detected experimentally. This is mainly due to the modal 
response of the structure which, for a given driving power, is 
drastically suppressed as its excitation level increases as we shall
prove in the study of a damped driven antenna done in 
section \ref{Secdampeddriven}. Moreover, the antenna is generally driven
by a harmonic force that is constant over the beam which  
does not allow for the observation of antisymmetric modes, 
$Y_{n}(u)$ with $n$ even. In what follows, we show how
to evaluate the frequency of the fundamental mode of the structure and 
explain the clustering phenomenon observed at both edges of each band.

\subsection{Small $\gamma$ solution, fundamental mode and mass loading} 
\label{smallgsec}

If a small solution to Eq. \eqref{secEq} exists, $\gamma \ll 1$, 
we can obtain its approximate expression by expanding Eq. \eqref{secEq}
around $\gamma=0$. In this case we find that
\begin{equation} \label{secEqgamma0}
\gamma \simeq \left(\frac{\mu_b}{\mu_a}\right)^{1/4}\frac{l \beta}{L}.
\end{equation}
where $\mu_a = (m_b+2N m_t)/L$ is the mass per unit length of the entire
antenna (beam plus cantilevers).   
Replacing this result in Eq. \eqref{Omega} and using the fact that
$(\mu_t {\cal E}_b)/(\mu_b {\cal E}_t)=1$ (see Eqs. \eqref{EtvsEb} and  
\eqref{mutvsmub}) yields
\begin{equation} \label{Omegasmallgamma}
\omega \simeq \sqrt{\frac{{\cal E}_b}{\mu_a}}
\left(\frac{\beta}{L}\right)^2.
\end{equation}
In this limit, the inertia of the cantilevers is negligible 
and the frequency of the vibrational mode is the frequency of a 
clamped-clamped beam whose mass per unit length includes the mass of the 
central beam plus the mass of the cantilevers.
According to Eqs. \eqref{Huvsol} and \eqref{Aju}, 
the mode shape  $H(u,v)$ corresponding 
to this small $\gamma$ solution is found to be 
\begin{equation} \label{Huvsmallgamma}
H(u,v) \simeq \frac{\gamma^4}{24}v^2(v^2-4v+6) Y(u).
\end{equation}  
Notice that this function is proportional to $\gamma^4$ and is 
very small. This confirms that the cantilevers hardly move at all and that 
their mass only loads the central beam. Finally,
given our parameters, the condition $\gamma \ll 1$ is typically satisfied 
for the first positive root of Eq. \eqref{secEq} when $\beta=\beta_1$. 
It is then valid for $\gamma_{1,1}$, i.e. for the fundamental mode. Indeed,
using Eq. \eqref{secEqgamma0} and the antenna parameters of section 
\ref{SecIIsub1}, we find $\gamma_{1,1}\simeq 0.1874150\dots$ while the exact
numerical value reads $\gamma_{1,1}=0.1874136\dots$, i.e. a relative error
of $7\times10^{-4}$ \%.

\subsection{Frequency clustering} \label{SecFreqClus}

The {\em frequency clustering} evoked in \ref{SecFreqSpec}, which has also
been observed numerically in 3D finite-element simulations,
can be explained with the help of secular equation \eqref{beta4}. 
To evaluate the accumulation of frequencies occurring at the upper 
edge of the $k$th
band ($\omega_{\infty,k}$) and at the lower edge of the $(k+1)$th band, we seek
a perturbative solution of Eq. \eqref{beta4} around $\gamma \sim 
\gamma_{\infty,k}$ given in Eq. \eqref{seccanti}. 
Using Eq. \eqref{secEq} and $\gamma_{n,\tilde{k}} 
\simeq \gamma_{\infty,k} - 
Q(\gamma_{\infty,k},\beta_n)/Q'(\gamma_{\infty,k},\beta_n)$, 
where $\tilde{k}$ stands either for $k$ or $k+1$, and $Q'$ is the derivative 
of $Q$ with respect to $\gamma$, we obtain
\begin{multline} \label{gammaapp}
\hskip-2ex \gamma_{n,\tilde{k}} \simeq \gamma_{\infty,k}+ 
\left\{\!
\frac{L\gamma_{\infty,k}}{2\alpha N l}
\left(\frac{t_{\gamma_{\infty,k}} \!-\! th_{\gamma_{\infty,k}}}
{t_{\gamma_{\infty,k}} \!+\! th_{\gamma_{\infty,k}}}\right)\!\!
\left[1\!-\!\left(\frac{l\beta_n}{L\gamma_{\infty,k}}\right)^4\right]
\right. \\
\left.
+ th_{\gamma_{\infty,k}} -\frac{2}{t_{\gamma_{\infty,k}} + 
th_{\gamma_{\infty,k}}} + \frac{1}{\gamma_{\infty,k}}
\right\}^{-1}\, ,
\end{multline} 
where $t_{\gamma}=\tan \gamma$ and $th_{\gamma}=\tanh \gamma$. This solution
is valid as far as the corrective term is very small compared to
$\gamma_{\infty,k}$. This leads to the two conditions below. 
\subsubsection{Upper band edge}
With respect to the frequency clustering at the upper band edge of the $k$th 
band, Eq. \eqref{gammaapp} starts to be valid as soon as $\beta_n$ is large 
enough. In this case, it is easy to see that this approximately leads to
the condition
\begin{equation} \label{condUBE}
\beta_n \gg \frac{L\gamma_{\infty,k}}{l}\left(1+
\frac{2\alpha N l}{L}\right)^{1/4}.
\end{equation}
This condition can always be satisfied for large enough $n$, 
given that in this 
case, $\beta_n \simeq (n+1/2)\pi$ (for simplicity, we have discarded
the term $(t_{\gamma_{\infty,k}} - th_{\gamma_{\infty,k}})/(
t_{\gamma_{\infty,k}} + th_{\gamma_{\infty,k}})$ that is close to unity). The 
longer the cantilever, the easier the condition \eqref{condUBE} to be 
satisfied. 
For an antenna structure with high cantilever to beam length 
ratio, we may expect to observe
a clustering of modes around the frequency determined in \eqref{Omega_infty}.
Experimentally, however, this very much depends on the sensitivity of the 
measuring device given that highly excited modes for the beam are difficult
to detect. Moreover, the range of validity of our model, determined by 
the condition $n<N$, restricts the possibility of observing
the beginning of an upper edge clustering to the fundamental band.

Provided \eqref{condUBE} is satisfied, Eq. \eqref{gammaapp} simplifies to
\begin{eqnarray}
\gamma_{n,k} - \gamma_{\infty,k} &\simeq &
\frac{NR \gamma^3_{\infty,k}}{\beta^4_n}\left(
\frac{th_{\gamma_{\infty,k}}+t_{\gamma_{\infty,k}}}
{th_{\gamma_{\infty,k}}-t_{\gamma_{\infty,k}}}\right) \nonumber \\
&=& - \frac{NR \gamma^3_{\infty,k}}{\beta^4_n}\left(
\frac{th_{\gamma_{\infty,k}}+t_{\gamma_{\infty,k}}}
{th_{\gamma_{\infty,k}} t_{\gamma_{\infty,k}}}\right)^2. 
\end{eqnarray} 
The second equality has been obtained from \eqref{seccanti}. As 
we can see, the corrective term to
$\gamma_{\infty,k}$ is always negative which
confirms that the frequencies are indeed accumulating at the upper edge 
of the band as $n$ increases.
It is also possible to show that the (adequately 
normalized) modal shape of the cantilever corresponding to $\omega_{n,k}$
factorizes asymptotically as the product of the $k$th mode 
of a simple cantilever by the $n$th mode of a clamped-clamped beam, 
$H_{n,k}(u,v) \sim \phi_k(v)Y_n(u)$ as $n \rightarrow \infty$. 
Thus, as the excitation of the beam becomes high enough, 
the modal shape of the antenna is given by a simple 
cantilever mode modulated by the profile of the central beam. The same result
is obtained for a weakly dissipative driven antenna in \ref{SecModalExpansion}.
\subsubsection{Lower band edge}
A frequency clustering at the lower edge of each band does not always occur.
This essentially depends on the geometry of the antenna. For the correction
given in Eq. \eqref{gammaapp} to be valid, 
\begin{equation} \label{condLBE}
\beta_n \ll \frac{L\gamma_{\infty,k}}{l}\left[1-
\frac{2\alpha N l}{L}\left(
\frac{th_{\gamma_{\infty,k}}+t_{\gamma_{\infty,k}}}
{th_{\gamma_{\infty,k}}-t_{\gamma_{\infty,k}}}\right)\right]^{1/4}
\end{equation}
has to be satisfied. For large cantilever to beam length ratio, however,
inequality \eqref{condLBE} cannot be satisfied because the square-bracketed 
term becomes negative. This means either that 
expanding the secular equation around the upper edge of the $k$th band
does not provide any reliable information on the lower edge of the $(k+1)$th
band or simply that no clustering occurs at the lower band edge. For the 
geometry of the antenna described in the introduction, frequency clustering 
occurs and the band gap $\Delta_{k,k+1}$ 
between the $k$th and $(k+1)$th bands can be accurately 
evaluated with the help of Eq. \eqref{gammaapp} as
\begin{equation} \label{bandgap}
\Delta_{k,k+1} \equiv \gamma_{1,k+1} - \gamma_{\infty,k}.
\end{equation}
This expression is accurate to less than 5\% for the first gap and to less 
than 0.5\% for the second one and its accuracy improves drastically 
as $k$ increases.  

\subsection{Energy and Lagrangian}

\subsubsection{Modal energy}

Using expressions \eqref{H2} and (\ref{modeyeta}-\ref{Eqbeta}), 
we can evaluate the energy $E_{n,k}$ of the mode 
$\varphi_{n,k}=(y_n(x,t),\eta_{n,k}(x,\xi,t))^T$. 
After some algebra, we find
\begin{equation} \label{Energynk}
E_{n,k} = \frac{M_{n,k} A^2 \omega^2_{n,k}}{2}, 
\end{equation}
where
\begin{equation} \label{Mnk}
M_{n,k}=m_b+2N m_t L_{n,k}. 
\end{equation}
The quantity $L_{n,k}$, that takes into account the dynamics 
of the cantilevers at the level of the effective mass $M_{n,k}$, is given by 
$L_{n,k}=L(\gamma_{n,k})$ where, defining $h(v) \equiv \tilde{H}(v)+1$, 
\begin{equation} \label{Lnk}
L(\gamma) = \int_0^1 h^2(v) \, dv = 
\frac{1}{4}\left(\frac{\cos \gamma +\cosh \gamma}
{1+\cos \gamma \cosh \gamma}\right)^2+
\frac{3}{2}\frac{A_2(\gamma)}{\gamma}. 
\end{equation}
Incidentally, we also note a relation that will prove useful when dealing with
the motion of a driven weakly damped structure in section \ref{secWeakDiss}
\begin{equation} \label{Lnkbis}
L(\gamma) = \frac{1}{2\gamma^3}\frac{d}{d\gamma}
\left(\gamma^3A_2(\gamma)\right).
\end{equation}
Eq. \eqref{Energynk} is exactly the energy of an effective harmonic 
oscillator
with amplitude $A$, frequency $\omega_{n,k}$ and mass $M_{n,k}$. 
Notice that, according to our definition of $Y(u)$ in \eqref{Yusol},
the amplitude is defined as $A^2 = \int_0^1 \! Y_n^2(u)\, du$
which is independent of $n$ and $k$.
Other definitions of the amplitude (like the maximal deflection of the central 
beam at the mid point, for example) are typically mode-dependent and 
redefine the effective mass of the structure.
As we see, the effective
mass of the antenna is renormalized by the factor $L_{n,k}$ that affects 
the total mass of the cantilevers, $2N m_t$. In the specific case of 
the fundamental mode, we have shown in the previous section that 
$\gamma_{1,1} \ll 1$. Therefore, 
$L_{1,1}\simeq 1$ and the effective mass of the antenna, $M_{1,1}\simeq 
m_b+2N m_t$, is its actual mass. This confirms the finding of section 
\ref{smallgsec} that in the 
fundamental mode the cantilevers have a pure mass loading effect.
We can also use Eq. \eqref{Energynk} to define an effective mode-dependent 
spring constant $\kappa_{n,k}$ as
\begin{equation} \label{kappank}
\kappa_{n,k} = M_{n,k} \omega^2_{n,k}. 
\end{equation}
Once again, this result depends on the definition chosen for the amplitude
and is valid here for $A^2 = \int_0^1 \! Y_n^2(u)\, du$.

\subsubsection{Energy partition}

\begin{figure}
\centering
\includegraphics[width=.47\textwidth]{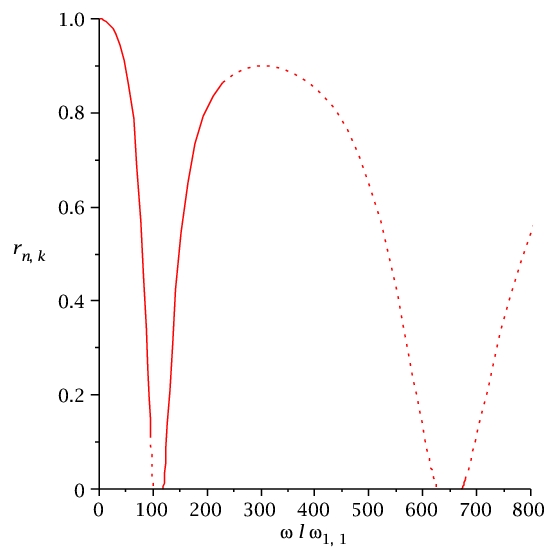}
\caption{Ratio, $r_{n,k}$ (see Eq. \eqref{rnk}), of the maximal 
elastic energy of the beam to the total energy of the antenna versus 
the normalized frequency, $\omega_{n,k}/\omega_{1,1}$. Antenna 
parameters are given in section \ref{SecIIsub1} and $N=20$. 
Solid lines indicate physically relevant results for the continuum model 
($n<N$) and dotted lines the results for $(n>N)$.}\label{Fig_rnkvsfnk}
\end{figure}

To complete the modal analysis of the antenna, we evaluate the
energy partition of each mode that is, we compare the energies 
of the cantilever continuum and of the central beam to understand
which part of the structure is the most active. For a mode, the elastic
energies of the central beam and the cantilever continuum reach their 
maximum at the same time and at this point their kinetic
energy is zero. The sum of the maximal elastic energies of the beam
and the continuum is then equal to the total energy of the system. 
We thus choose to calculate the ratio, $r_{n,k}$, 
of the maximal elastic energy of the beam, $U_{b;n,k}$, 
to the total energy of the structure, $E_{n,k}$, 
as a function of the frequency of the mode, 
$\omega_{n,k}$. The function $r_{n,k}$ ranges
from 0 to 1 and is expected to be useful in the design of antenna structures
with a specific task.
The central beam elastic energy is given by 
\begin{equation} \label{CBenergy}
U_{b}[y(x,t)] = \int\limits_0^L dx 
\left[\frac{{\cal E}_b}{2}\left(\frac{\partial^2 y}{\partial x^2}\right)^2 
\right].      
\end{equation}
Using expressions (\ref{modeyeta}-\ref{Eqbeta}), we obtain
\begin{eqnarray} \label{TACBenergy}
U_{b;n,k} &\equiv& \max_t U_{b}[Y_n(x/L)\cos(\omega_{n,k}t)] = 
\frac{{\cal E}_b}{2L^3}A^2 \beta_n^4 \nonumber \\
&=& Nm_tA^2\omega_{n,k}^2
\left(\frac{m_b}{2Nm_t}+2\frac{A_2(\gamma_{n,k})}{\gamma_{n,k}}\right).     
\end{eqnarray}
Using expression \eqref{Energynk} for the total energy $E_{n,k}$, we 
finally have 
\begin{equation} \label{rnk}
r_{n,k} \equiv \frac{U_{b;n,k}}{E_{n,k}} = 
\left(\frac{\frac{m_b}{2Nm_t}+2\frac{A_2(\gamma_{n,k})}
{\gamma_{n,k}}}{\frac{m_b}{2Nm_t}+L_{n,k}}\right),
\end{equation} 
where $L_{n,k}$ is given in Eq. \eqref{Lnk}. In Fig. \ref{Fig_rnkvsfnk}
we display $r_{n,k}$ versus the normalized frequency 
$\omega_{n,k}/\omega_{1,1}$. We observe that
at the edges of all the bands but the lower edge of the first one, 
$r_{n,k}$ almost vanishes. This 
means that most of the energy of the structure is located in the cantilever
continuum. By contrast, in the middle of the bands, most of the energy
is stored in the central beam. This confirms the results of the previous
sections showing that close to the band edges, the parameter
$\gamma$ almost satisfies the ``cantilever'' secular equation 
$1+\cos \gamma \cosh \gamma=0$, which clearly indicates that the spectrum
is mainly governed by the dynamics of the cantilever continuum 
in this region while in the ``middle'' of the bands, 
$\gamma_{n,k} \sim (l/L)\beta_n$ 
(see Eq. \eqref{gammamidbandasympt}), which makes the spectrum close to that 
of a clamped-clamped beam. 
The peculiarity of the first band lies in the fact 
that for $\gamma_{n,1}$ close enough to zero, $r_{n,1}\sim 1-
\gamma_{n,1}^4/[20(\sigma+1)]$ where $\sigma=m_b/2Nm_t$. The energy
of the very first modes and, in particular, the fundamental, is thus 
essentially stored in the beam. This confirms the results obtained 
in section \ref{smallgsec}.    
    
\subsubsection{Lagrangian}

It is also interesting to evaluate the Lagrangian ${\cal L}_{n,k}(t)$ 
of the system and in 
particular its time average that we will use in the next section to 
evaluate the effects of small nonlinearities and damping on the structure.
We find
\begin{equation} \label{Lagrangenk}
{\cal L}_{n,k}(t) = -\frac{M_{n,k} A^2 \omega^2_{n,k}}{2}\cos (2\omega_{n,k}t)
\ \ \Rightarrow \ \ \langle {\cal L}_{n,k}(t) \rangle = 0,
\end{equation}
where $\langle \dots \rangle$ denotes the time average. The modal Lagrangian 
of the linear continuum model is then similar to a harmonic oscillator's and
its time average is zero. This result will be used in the perturbative 
treatment of a weakly nonlinear and dissipative antenna in section 
\ref{NLN2F}. 

\subsection{Fundamental and first collective modes: comparison with 
finite-element results} \label{SecComparison}

In this section, we compare the results of our continuum model (CM) to those
obtained with a finite element method (FE) that treats the vibrations 
of the antenna within the frame of three-dimensional
elasticity theory.   
We are primarily interested in the fundamental and first collective modes,
that is in the first modes of the first and second band, respectively.
These modes are easy to observe experimentally. They 
are related to the parameters $\gamma_{1,1}$ and $\gamma_{1,2}$,
respectively.
Using Eq. \eqref{secEq} and the antenna parameters given
in \ref{SecIIsub1},
we find that the two first roots of $Q(\gamma,\beta_1)$ with
$\beta_1= 4.73004\dots$
are $\gamma_{1,1}=0.187413\dots$ and $\gamma_{1,2}=2.046440\dots$. If we use 
Eq. \eqref{secEqgamma0}, one finds $\gamma_{1,1}\simeq 0.187415$. It 
shows that this approximate solution is very reliable for the fundamental
frequency. The corresponding frequency, obtained from Eq. \eqref{Omega}, 
is $f_{1,1}= \omega_{1,1}/2\pi \simeq 24.7$ MHz, slightly higher than 
the frequency observed in simulations ($23.6$ MHz). This might be 
explained by the effective cross-sectional stiffness that we approximate for the two material layers. The frequency 
of the first collective mode calculated from $\gamma_{1,2}=2.046440$ is
$f_{1,2}\simeq 2.94$ GHz which is much higher than 
$f^{\rm FE}_{1,2}\simeq 1.51$ GHz, the frequency from finite element simulations.
A reason for this discrepancy is that all elements of the antenna 
structure are considered as one-dimensional in our model. Consequently,
their length is the only dimension taken into account in the dynamics
of the system. However, in the real structure, the central beam has 
a nonzero width, $W$, on the order of the cantilever length, $l$. For motions
of the cantilevers comparable to the beam's, the shear momentum they
exert on both of its sides becomes large enough to bend it {\em laterally} 
with respect to its mid-line, an occurrence indeed confirmed by our 
three-dimensional finite-element simulations. 
In first approximation, we can account for this effect by assigning to the 
cantilevers an effective {\em dynamical} length ranging 
from their actual length, $l$, for small amplitude motions to $l+W/2$, for 
large ones. If we carry out the substitution $l \rightarrow l+W/2$ in 
$Q(\gamma,\beta_1)$ and reevaluate 
$\gamma_{1,2}$, we find $\gamma_{1,2}= 2.085115$
and a frequency $f_{1,2}\simeq 1.55$ GHz, within 3 percent of the simulated
value. 
\begin{figure*} 
\includegraphics[scale=0.8]{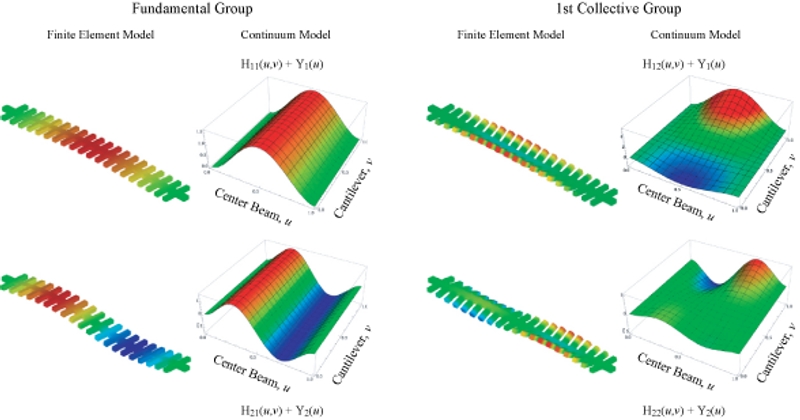} 
\caption{Comparison between finite element (FE) and continuum model (CM)
results for the fundamental mode (upper left panel), the first excited mode
of the fundamental group (lower left panel), the first collective mode 
(upper right panel), and the first excited collective mode (lower right panel).
The parameters for the antenna structure are those quoted in \ref{SecIIsub1}.} 
\label{Figmodeeta}
\end{figure*}

The shapes of the central beam predicted by the continuum model are 
the same for the fundamental and first collective modes. They are given by 
$Y_1(u)$, obtained from Eq. \eqref{Yusol} for 
$\beta=\beta_1$. The shapes of the cantilever 
continua, however, are different and given by equation \eqref{Huvsol} as
$H_{1,1}(u,v)$ and $H_{1,2}(u,v)$, respectively. A comparison of these results
to those of the finite element simulation is done in Fig. \ref{Figmodeeta}.
The FE and CM results for the fundamental 
and first collective modes are given in the upper left and 
upper right panel, respectively. As indicated, the FE result is 
to the right of the CM result.
In both cases, the color code (online) indicates the displacement (amplitude)
of the elements with respect to the clamps of the central beam 
(i.e. its extremities). For the CM results, only half of the absolute 
displacement of the cantilever continuum, $Y_1(u)+H_{1,k}(u,v)$, $k=1,2$, 
$v \geq 0$, is displayed. For $v=0$, this displacement is precisely the beam's,
$Y_1(u)$. As we see, the continuum model predictions are in excellent 
agreement with the finite element results. For both the fundamental and the 
first collective modes, the shape of the central beam is 
in the fundamental mode of a bare beam, $Y_1(u)$. 
In the fundamental mode, the deflection of the cantilever continuum with 
respect to the beam is imperceptible as predicted by
the continuum model in the small $\gamma$ limit (see \ref{smallgsec}). 
Indeed, from Eq. \eqref{Huvsmallgamma}, we 
have $H_{1,1}(u,v) \propto \gamma_{1,1}^4 Y_1(u)$, which means that the 
continuum deflection is roughly 5000 times smaller than that of the beam.  
This is markedly different for the first collective mode where the motion 
of the continuum is on the order of the beam amplitude as seen on the
upper right panel. Note that, if their amplitudes are very 
different, the actual shapes of the continua for the fundamental and
first collective modes are similar and are given to a very good approximation
by the fundamental mode of a bare cantilever. This somewhat surprising 
fact becomes clear on the modal expansion of the cantilever continuum,
$H_{n,k}(u,v)\equiv \tilde{H}_{n,k}(v)Y_n(u)$. Introducing the normalized 
cantilever modes, $\psi_l(v)$, satisfying 
$\psi^{(iv)}_l(v)=\gamma^4_{\infty,l}\psi_l(v)$, we obtain from Eq. 
\eqref{Huv2} 
\begin{equation} \label{Htildamodexp}
\tilde{H}_{n,k}(v) = \sum_{l=1}^{\infty} 
\frac{\gamma_{n,k}^4 \Psi_{l}}{\gamma_{\infty,l}^4-\gamma_{n,k}^4}
\psi_l(v),
\end{equation} 
where $\Psi_{l}=\int_0^1\psi_l(v)dv=-(2/\gamma)
(\cos \gamma+\cosh \gamma)/(\sin \gamma+\sinh \gamma)$,
where $\gamma=\gamma_{\infty,l}$. Now, we have seen in section 
\ref{SecFreqClus} that, for $k \geq 2$, $\gamma_{n,k}$ is
very close to $\gamma_{\infty,k-1}$. According to Eq. \eqref{Htildamodexp}, 
this means that 
the mode selected for the cantilever continuum is $\psi_{k-1}(v)$. In
particular, for the first collective mode, 
$\gamma_{1,2}$, the continuum adopts basically 
the shape of the {\em fundamental} cantilever mode, $\psi_1(v)$. 
For the modes of the fundamental band ($k=1$), 
this is also the fundamental cantilever mode
that is predominantly excited, even if its amplitude is so small compared
to the beam's that the cantilevers seem flat at the scale of the figure. 

To analyze further the similarities between the FE and CM results, 
the first excited modes of the fundamental and collective bands are displayed
in the lower left and right panels of Fig. \ref{Figmodeeta}, respectively. 
We clearly observe in both cases that the beam deflection is in the first
excited mode of a bare beam, $Y_2(u)$, as predicted by the continuum 
model. Moreover, as for the fundamental mode,
the cantilevers of the first excited mode of the fundamental group are hardly
moving. Evaluating the first root of $Q(\gamma,\beta_2)=0$ with 
$\beta_2= 7.85320\dots$ yields $\gamma_{2,1}=0.31114\dots$. This is small
enough for the results of section \ref{smallgsec} to hold. Indeed, expression
\eqref{secEqgamma0} gives $\gamma_{2,1}\simeq 0.31116\dots$. Therefore,
the continuum deflection can safely be evaluated from Eq. \eqref{Huvsmallgamma}
and we find that it is roughly 100 times smaller than the beam deflection, 
which explains that no cantilever motion can be detected in the FE results.
Once again, this is markedly different for the first excited mode of the 
collective group whose cantilevers, according to the finite element results, 
are experiencing a deflection comparable to the beam amplitude in full
agreement with the predictions of the continuum model. Note that for this two 
first excited modes of the fundamental and collective bands, the cantilever
deflections adopt the shape of the fundamental mode of a bare cantilever, 
$\psi_1(v)$, as indicated by the modal expansion \eqref{Htildamodexp}. 

Even though one-dimensional in essence, the continuum model gives a good
qualitative understanding of the modes of the 3D antenna. It correctly 
reproduces the modal shapes of the central beam and the cantilevers 
observed in 
the finite element simulation and is able to explain the frequency
clustering occurring in the spectrum of the structure. We use it in the next 
section as the basic model in the investigation 
of the effect of a two-frequency 
driving on the response of a weakly nonlinear and dissipative antenna.

\subsection{Driven damped system: Exact solution} \label{Secdampeddriven}

We conclude this section devoted to the linear system 
by calculating the exact solution of a damped antenna structure
driven by a spatially uniform harmonic force density, $f(t)$, in
the continuum approximation. 
This force density can, for instance, be exerted by an external 
magnetic field, $B$, 
orthogonal to the flexural vibrations of the central beam, acting on an 
ac-current with frequency $\omega_d$ passed through the thin layer 
of gold coating the structure. 
The beam vibrations then generate in turn an electromotive voltage $V_{emf}(t)$
at the clamped ends of the antenna that is proportional to the
rate of change of the magnetic flux,
\begin{equation} \label{Vemf}
V_{emf}(t) = B \int\limits_0^L \frac{\partial y(x,t)}{\partial t}\, dx.
\end{equation}   
Ultimately, this voltage can be monitored to determine 
the motion of the central beam of the structure and to derive its 
``spectrum'' i.e., the time root-mean-square of the induced voltage, 
$\sqrt{\langle V^2_{emf}(t)\rangle}$, versus the driving frequency, 
$\omega_d$. This detection scheme has been used in the previously reported experimental measurements of the antenna resonators in Ref. 
\cite{gaidarzhy-prl,imboden}. 
Other activation/detection schemes such as electrostatic, piezo-electric, and optical are typically also sensitive to the average transverse displacement of resonating elements, and the following analysis applies quite generally.

\subsubsection{Model}
   
Supplementing the equations of motion \eqref{y2}-\eqref{eta2}
with damping and driving terms we obtain
\begin{eqnarray}
&& \hskip-5ex {\cal E}_b \frac{\partial^4 y}{\partial x^4} + \mu_b 
\frac{\partial^2 y}{\partial t^2} + \nu_b 
\frac{\partial y}{\partial t} = - \frac{2 {\cal E}_t N}{L}
\left. \frac{\partial^3 \eta}{\partial \xi^3} 
\right|_{\xi=0} + f(t) \label{y2bis}\\
&& \hskip-5ex {\cal E}_t \frac{\partial^4 \eta}{\partial \xi^4} + \mu_t 
\frac{\partial^2 \eta}{\partial t^2}+ \nu_t 
\frac{\partial \eta}{\partial t} = -\mu_t 
\frac{\partial^2 y}{\partial t^2}-\nu_t 
\frac{\partial y}{\partial t}. \label{eta2bis} 
\end{eqnarray}
Boundary conditions for the beam and cantilever deflections 
are the same as in \eqref{bcy2} and \eqref{bceta2}.
The harmonic force density is given by $f(t)=f\cos (\omega_d t)$. Note that 
in Eq. \eqref{eta2bis}, the damping term affecting the cantilevers 
involves their absolute displacement, $y(x,t)+\eta(x,\xi,t)$, rather than 
their relative displacement, $\eta(x,\xi,t)$.
The choice of an appropriate damping term depends of course 
on the type of damping experienced by the structure. 
To simplify, we consider here that damping occurs through 
air friction and is then proportional to the absolute velocity of the 
cantilevers. Moreover, it is proportional to the surface in contact 
with the ambient air. For that reason, the damping per unit length is 
proportional to the width of the element involved and then 
$\nu_t/\nu_b=w/W=\mu_t/\mu_b={\cal E}_t/{\cal E}_b$.  
Material damping, whether of the viscoelastic or hysteretic type 
(see for instance Ref. \cite{Silva}), would essentially affect the rigidities 
of Eqs. \eqref{y2bis} and \eqref{eta2bis} in such a way that 
${\cal E}$ is replaced by ${\cal E}+{\cal E}^* \partial/\partial t$. 
The exact solution of system 
\eqref{y2bis}-\eqref{eta2bis} can still be obtained in this case.

\subsubsection{Exact solution}

Introducing the Fourier transform $\tilde{g}(\omega) = \int
dt \, e^{i\omega t} g(t)$ and its inverse $g(t) = 
\int d\omega \, e^{-i\omega t} \tilde{g}(\omega)/(2\pi)$
and using the dimensionless quantities defined in \eqref{nondim}, we can cast
Eqs. \eqref{y2bis}-\eqref{eta2bis} into 
\begin{eqnarray}
&& \hskip-5ex \frac{\partial^4 \tilde{y}}{\partial u^4} - 
\mu_c^4 \tilde{y} = - N R 
\left. \frac{\partial^3 \eta}{\partial v^3} 
\right|_{v=0} \!\!+\! \tilde{F}(\omega), \label{y2Fourier}\\
&& \hskip-5ex \frac{\partial^4 \tilde{\eta}}{\partial v^4} -\gamma_c^4 
\tilde{\eta} = \gamma_c^4 \tilde{y}, \label{eta2Fourier} 
\end{eqnarray}
where $F(t)=F \cos (\omega_d t)$ with $F= L^4 f/{\cal E}_b$ and 
\begin{eqnarray} \label{Complexgammaandmu}
\gamma_c^4 &=& \frac{l^4}{{\cal E}_t}(\mu_t \omega^2 + i\nu_t \omega), 
\nonumber \\
\mu_c^4 &=& \frac{L^4}{{\cal E}_b}(\mu_b \omega^2 + i\nu_b \omega)=
\left(\frac{L\gamma_c}{l}\right)^4.
\end{eqnarray}
These last parameters are the complex version (because of the presence
of damping) of the parameters $\mu$ and $\gamma$ defined in \eqref{nondim}.
Note also that $F$ has the dimension of a length.
As in the earlier case, we can solve Eq. \eqref{eta2Fourier} with its 
boundary conditions:
\begin{equation} \label{etasolFourier}
\tilde{\eta}(u,v,\omega) = \tilde{H}(v,\omega) \tilde{y}(u,\omega),
\end{equation}
where 
\begin{multline} \label{HvomegaFourier}
\tilde{H}(v,\omega)=\left[A_1(\gamma_c) \cos (\gamma_c v) + 
A_2(\gamma_c) \sin (\gamma_c v) \right. \\ 
\left. + A_3(\gamma_c) \cosh (\gamma_c v) + 
A_4(\gamma_c) \sinh (\gamma_c v) -1 \right] .
\end{multline}
The coefficients $A_i$ are the same as those given in \eqref{Aju}. 
Reinstating this expression in Eq. \eqref{y2Fourier}, we finally obtain
\begin{equation} \label{y2Fourierbis}
\frac{\partial^4 \tilde{y}(u,\omega)}{\partial u^4} - 
\beta_c^4 \tilde{y}(u,\omega) = \tilde{F}(\omega),
\end{equation}
where 
\begin{equation} \label{betacomplex}
\beta_c^4 = \mu_c^4 + \gamma_c^3 R N \frac{\cos \gamma_c \sinh \gamma_c + 
\sin \gamma_c \cosh \gamma_c}{1+\cos \gamma_c \cosh \gamma_c},
\end{equation}
which is the complex analog of the secular equation \eqref{beta4}. 
Finally, applying the appropriate boundary conditions to 
$\tilde{y}(u,\omega)$ we can, 
after some algebra, cast the solution to Eq. \eqref{y2Fourierbis} into
\begin{multline} \label{y2solFourier}
\tilde{y}(u,\omega) = \left\{ T(\beta_c)\left[\frac{\cos (\beta_c(u-1/2))}{\sin (\beta_c/2)}+\right. \right.\\
\left.\left. +\frac{\cosh (\beta_c(u-1/2))}{\sinh (\beta_c/2)}\right]-1\right\}
\frac{\tilde{F}(\omega)}{\beta^4_c},
\end{multline}
where we have defined
\begin{equation} \label{Tbetac}
T(\beta_c) = \frac{\tan (\beta_c/2)\tanh (\beta_c/2)}
{\tan (\beta_c/2)+\tanh (\beta_c/2)}.
\end{equation}
Now, the driving $f(t)$ being harmonic, we have 
$\tilde{F}(\omega)=F\pi[\delta(\omega-\omega_d)+
\delta(\omega+\omega_d)]$. Noticing that the sign inversion
$\omega \rightarrow -\omega$ amounts to taking the complex conjugate, we can
finally show: 
\begin{equation} \label{yxtRe}
y(x,t) = F {\rm Re} \{\exp (-i\omega_d t) \tilde{y}_h(u,\omega_d)\},
\end{equation}
 where $\tilde{y}_h(u,\omega)=\tilde{y}(u,\omega)/\tilde{F}(\omega)$. Then 
\begin{multline} \label{y2solFourierback}
y(x,t) = F {\rm Re} \left\{ \frac{e^{-i\omega_d t}}{\beta^4_c} 
\left[ T(\beta_c)\left(\frac{\cos (\beta_c(u-1/2))}{\sin (\beta_c/2)}+ 
\right. \right. \right.\\ + \left. \left.\left.
\frac{\cosh (\beta_c(u-1/2))}{\sinh (\beta_c/2)}\right)-1\right]\right\}.
\end{multline}
Expression \eqref{y2solFourier} makes it clear that the shape of the 
central beam induced by the force density $f(t)$ is symmetric with respect
to its midpoint, $u=1/2$. 
This is expected as the force density itself possesses this symmetry. 
Consequently,
none of the antisymmetric modes of the central beam are excited 
by this method. Moreover, in presence 
of dissipation, the beam shape never corresponds to an exact symmetric 
modal shape even when the driving frequency is one of the structure modal 
frequencies. We can see that for weak dissipation, however, the 
denominator of $T(\beta_c)$ becomes small (on the order of $\nu_t$)
when $\omega_d \sim \omega_{2n+1,k}$. This is because 
the solutions to $\tan (x/2)+\tanh(x/2)=0$ are precisely the $\beta_{2n+1}$'s. 
This eventually leads $y(x,t)$ to assume a shape close to the mode 
$Y_{2n+1}(u)$.

\subsubsection{Modal expansion} \label{SecModalExpansion}

As we have seen earlier, the modes of the antenna
structure are such that the central beam possesses the {\em exact} shape 
of a clamped-clamped mode, $Y_n(u)$ (see Eq. \eqref{Yusol}). However, the 
force $f(t)$ applied to the beam excites now {\em all} symmetric 
clamped-clamped modes. To get a sense of which modes are predominantly
excited, we first expand the Fourier transform 
of the deflection as $\tilde{y}(u,\omega)=\sum_n \tilde{y}_n(\omega) Y_n(u)$, 
insert it in Eq. \eqref{y2Fourierbis} and solve for $\tilde{y}_n(\omega)$. 
The beam deflection given by Eq. \eqref{yxtRe} then reads 
\begin{equation} \label{y2solmodes}
y(x,t) = F \sum_{n=0}^{\infty} \frac{\Gamma_{2n+1}\cos \left(\omega_d t+\vartheta_{2n+1}(\omega_d)\right)}{\left|\beta_c^4(\omega_d)-\beta_{2n+1}^4\right|}
Y_{2n+1}(u),
\end{equation} 
where $\Gamma_{n}=\int_0^1 Y_n(u)du$ and where $\vartheta_{n}(\omega) =\arg 
(\beta_c^4(\omega)-\beta_{n}^4)$. The reason why the sum runs exclusively 
over odd numbers in Eq. \eqref{y2solmodes} is because integrals 
of antisymmetric clamped-clamped modes, $\Gamma_{2n}$, vanish which confirms
that none of them is excited by $f(t)$. For symmetric modes, a 
simple calculation yields 
$\Gamma_{2n+1}=4\tan (\beta_{2n+1}/2)/\beta_{2n+1}$. Clearly, expression 
\eqref{y2solmodes} shows that the mode $Y_{2n+1}(u)$ is singled out 
when the driving frequency is close to one of the modal frequencies 
$\omega_{2n+1,k}$ and the dissipation is weak enough  
($|\beta_c^4(\omega_d)-\beta_{2n+1}^4| \propto \nu_t$ in this case, see 
\ref{secWeakDiss}).

Also of interest is the modal expansion of the cantilever continuum. 
Introducing the normalized cantilever modes, $\psi_k(v)$, $k\geq 1$, 
satisfying $\psi^{(iv)}_k(v)=\gamma^4_{\infty,k}\psi_k(v)$, we can expand 
$\tilde{H}(v,\omega)$ in \eqref{HvomegaFourier} as 
$\tilde{H}(v,\omega)=\sum_k \tilde{H}_k(\omega) \psi_k(v)$, uses Eq. 
\eqref{etasolFourier} and insert it in Eq. \eqref{eta2Fourier}, 
and finally solve for $\tilde{H}_k(\omega)$. This yields
\begin{equation} \label{eta2solmodes}
\tilde{H}(v,\omega) = \sum_{k=1}^{\infty} 
\frac{\gamma_c^4(\omega) \Psi_{k}}{\gamma_{\infty,k}^4-\gamma_c^4(\omega)}
\psi_k(v),
\end{equation} 
where $\Psi_{k}=\int_0^1\psi_k(v)dv=-(2/\gamma)
(\cos \gamma+\cosh \gamma)/(\sin \gamma+\sinh \gamma)$,
where $\gamma=\gamma_{\infty,k}$. Hence, $\Psi_{k}\simeq -2/\gamma_{\infty,k}$,
as $k$ becomes large. The interesting point about this calculation is to show 
what cantilever mode is selected according to the driving frequency. When 
the latter is close to the modal frequency $\omega_{n,k}$ with $n$ small, and
provided the dissipation is weak enough, $\gamma_c(\omega_{n,k})\simeq 
\gamma_{n,k}$. Now, we have seen in section \ref{SecFreqClus} that, 
for $k \geq 2$, $\gamma_{n,k}$ is
very close to $\gamma_{\infty,k-1}$. According to Eq. \eqref{eta2solmodes}, 
this means that for $\omega_d \sim \omega_{n,k}$, 
the mode selected for the cantilever continuum is $\psi_{k-1}(v)$. In
particular, if the system is driven near the first 
{\em collective} mode frequency, 
$\omega_{1,2}$, the continuum adopts basically the shape of the {\em 
fundamental} cantilever mode. Note that, for driving frequencies in the 
fundamental band ($k=1$), this is always the fundamental cantilever mode
that is predominantly excited. For other bands $(k\geq 2)$, the continuum
interpolates between the shape of the mode $\psi_{k-1}(v)$ close to the lower 
band edge and $\psi_{k}(v)$ close to the upper band edge. In practice, however,
as our model provides consistent results for $n<N$ only, the range of 
physically relevant frequencies is restricted to the lower band edge. For all
peaks but those of the fundamental band, therefore, the relevant modes of 
the structure at frequency $\omega_{n,k}$ are $Y_n(u)$ and $\psi_{k-1}(v)$.

\subsubsection{Amplitude-frequency spectrum}

Most of the time, this is not the beam deflection $y(x,t)$ but rather its
average over the beam length (or its time-derivative, see below) 
that is detected. For that reason, we define 
\begin{equation} \label{y2solaverage}
\bar{y}(t) \equiv \frac{1}{L}\int\limits_0^L y(x,t) dx. 
\end{equation}
Using expression \eqref{y2solFourierback}, we immediately obtain
\begin{equation} \label{y2solaverageexact}
\bar{y}(t) = {\cal A}(\omega_d) \cos (\omega_d t - \theta(\omega_d)),
\end{equation}
where the amplitude ${\cal A}(\omega)$ and the phase $\theta(\omega)$ are 
respectively given by 
\begin{eqnarray} \label{y2averageampph}
{\cal A}(\omega) &=& \frac{F}{|\beta_{c}^4|} \left|\frac{4T(\beta_{c})}
{\beta_{c}}-1\right|, \nonumber \\ 
\theta(\omega) &=& \arg \left(\frac{1}{\beta_{c}^4} 
\left[\frac{4T(\beta_{c})}{\beta_{c}}-1\right]\right).
\end{eqnarray} 
\begin{figure*}
     \centering
     \subfigure[]{
          \label{Fig_spectrum_driven}
          \includegraphics[height=.47\textwidth,width=.47\textwidth]{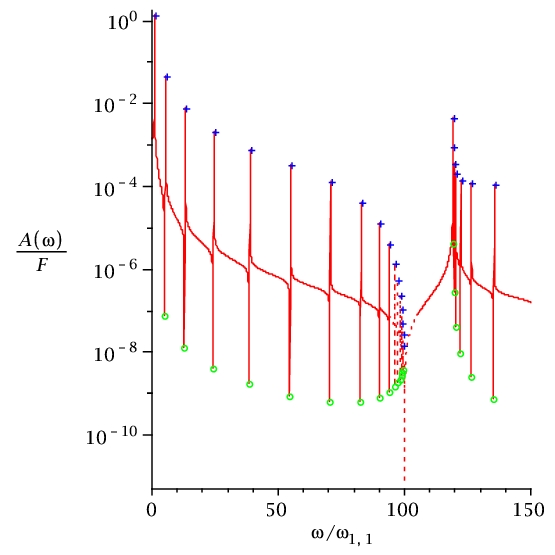}}
     \subfigure[]{
          \label{Fig_spectrum_driven_zoom}
          \includegraphics[width=.47\textwidth]{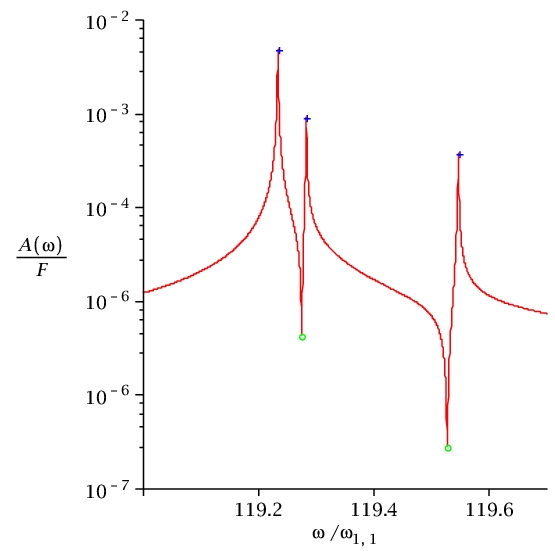}}
     \caption{(a): Amplitude-Frequency spectrum, (${\cal A}(\omega)/F$ versus 
$\omega/\omega_{1,1}$), obtained from Eq. \eqref{y2averageampph} 
for $N=20$ and the antenna parameters given in \ref{SecIIsub1}. Solid lines 
correspond to $n<N$ (physically relevant frequencies) and the dotted one to 
$n>N$.  (b): Zoom in the region of the 1st collective frequency
peak. For both (a) and (b), analytical results for peaks $(+)$ and dips 
$(\circ)$ are obtained from Eqs. \eqref{Apeaknk} and \eqref{Adipnk}, 
respectively.}
     \label{Fig_Specdriven}
\end{figure*}
Fig. \ref{Fig_spectrum_driven} displays the amplitude-frequency spectrum 
for the antenna parameters given in \ref{SecIIsub1}. Quantities plotted 
are the dimensionless amplitude, ${\cal A}(\omega)/F$, given by Eq. 
\eqref{y2averageampph}, versus the dimensionless
frequency $\omega/\omega_{1,1}$. Therefore, the peak of the fundamental 
is located at 1. The peak of the first collective mode (first frequency of 
the second band) is approximately located at $\omega_{1,2}/\omega_{1,1}=119.23$
as seen on Fig. \ref{Fig_spectrum_driven_zoom}.  
The dissipation parameter has been chosen in such a way as to provide
a fundamental peak with quality factor $Q = \mu_t \omega_{1,1}/\nu_t$
on the order of $\sim 10^3$, a typical value in experiments. The 
physically relevant
part of the spectrum (corresponding to $n<N$, that is 
$|\beta_c(\omega)|<(N+1/2)\pi$), is drawn in solid line while 
the irrelevant part is drawn in dotted line. Because of the weak damping,
frequency peaks occur 
when the driving frequency is close to one of the modal frequencies, 
$\omega_{2n+1,k}$. As seen in Fig. \ref{Fig_spectrum_driven}, peak 
amplitudes decrease rapidly as the frequency increases within a given band. 
But the first peak of the collective band, approximately located at
$\omega_{2,1}\simeq 119.23 \, \omega_{1,1}$, is much higher than the last peaks
of the fundamental band. That is why it is easily observed in experiments.
An analytical formula for peak heights is obtained in the next section
in the weakly dissipative regime. 

Also of interest are the dips occurring slightly
before the peaks. Their occurrence is due to the fact that at certain
frequencies, the integral of the shape of the central beam over its 
length is close to zero. This is especially important for detection 
schemes involving an electromotive voltage because the signal is close
to zero in this case. For a {\em non-dissipative system}, 
there are frequencies such that the integral is exactly zero. 
The shape of the central beam resembles the modal shapes 
$Y_{2n+1}(u)$, $n\geq 1$, in this case. 
From Eq. \eqref{y2averageampph},
we see that spectral dips are given by the simple relation $4T(\beta)=\beta$,
where $T(\beta)$ is given in \eqref{Tbetac}. Denoting with a hat all 
quantities related to the dips, we can show that the solution of the 
previous equation reads approximately 
\begin{equation} \label{betadip}
\hat{\beta}_{2n+1} \simeq \beta_{2n+1}+\frac{4}{4-\beta_{2n+1}}, \ n\geq 1.
\end{equation}
This explains why dips are close to the frequency peaks.
The corresponding frequencies are obtained by solving
the secular equation $Q(\gamma,\hat{\beta}_{2n+1})=0$ for $\gamma$ 
(see Eq. \eqref{secEq}) whose solutions $\hat{\gamma}_{2n+1,k}$ define 
$\hat{\omega}_{2n+1,k}=\sqrt{{\cal E}_t/\mu_t}(\hat{\gamma}_{2n+1,k}/l)^2$.
Last, the deepest dip occurring around $100$ in
Fig. \ref{Fig_spectrum_driven} is due to the upper band edge of the 
fundamental band, obtained in the non dissipative case for 
$\omega_{\infty,1}/\omega_{1,1}=(\gamma_{\infty,1}/\gamma_{1,1})^2
\simeq 100.103$. Though, this part of the spectrum is not physically relevant.

\subsubsection{Weak dissipation} \label{secWeakDiss}

Analytical results for weakly dissipative systems are obtained 
by expanding the quantities of interest around their value in absence of 
dissipation $(\nu_t=0)$. Let us assume that, for $\nu_t=0$,  
$\omega_0$ is the frequency of interest, a modal or dip frequency, 
for example. Then $\gamma_0=l(\mu_t\omega_0^2/{\cal E}_t)^{1/4}$ and 
$\beta_0$ given by \eqref{beta4} are real. For $\omega \sim \omega_0$, we have 
$\beta_c^4(\omega,\nu_t)\simeq \beta_0^4+[(\omega-\omega_0)
\left.\partial \gamma_c^4/\partial \omega\right|_0+
\nu_t \left.\partial \gamma_c^4/\partial \nu_t\right|_0]
\left.\partial \beta_c^4/\partial \gamma_c^4\right|_0$. Now, using Eq. 
\eqref{betacomplex} for $\beta_c$, Eq. \eqref{Complexgammaandmu} 
for $\gamma_c$ and the result obtained in \eqref{Lnkbis} yields
\begin{multline} \label{beta4complexexpand}
\left|\beta_c^4(\omega)-\beta_0^4\right| \simeq \frac{L^4\mu_t}{{\cal E}_t}
\left[1+\frac{2Nm_t}{m_b}L(\gamma_0)\right]\times \\
\times \sqrt{4\omega_0^2\left(\omega-\omega_0\right)^2+
\left(\frac{\nu_t\omega_0}{\mu_t}\right)^2},\ (\omega \sim \omega_0) 
\end{multline}
Notice that, because $\omega \sim \omega_0$, 
the function under the square root is, in first order in $\nu_t$, 
equivalent to $(\omega^2-\omega_0^2)^2+(\nu_t\omega/\mu_t)^2$, which is
the typical form for a harmonic oscillator. But $\beta_c(\omega,\nu_t)$ 
has to be expanded up to second order in $\nu_t$ and $\omega-\omega_0$
to compute the correct frequency shift induced by the damping 
because the latter is of order $\nu_t^2$.  

By expanding $\beta_c(\omega)$ around $\beta_{2n+1}$ and using Eq.
\eqref{y2averageampph}, we find that the 
amplitude of the peak with frequency $\omega_{2n+1,k}$ is given by 
${\cal A}(\omega)\simeq 16 F \tanh^2(\beta_{2n+1}/2)/(\beta_{2n+1}^2
|\beta_c^4(\omega)-\beta_{2n+1}^4|)$. Therefore, from Eq. 
\eqref{beta4complexexpand} we obtain the peak height as
\begin{equation} \label{Apeaknk}
{\cal A}^{\rm peak}_{m,k} \simeq \frac{16 {\cal E}_t F}{L^4 \nu_t}
\left[1\!+\!\frac{2Nm_t}{m_b}L(\gamma_{m,k})\right]^{-1}
\frac{\tanh^2 \left(\frac{\beta_{m}}{2}\right)}
{\beta^2_{m}\omega_{m,k}}.
\end{equation} 
where $m=2n+1$, $n\geq 0$.
In the same way, expanding $\beta_c(\omega)$ around $\hat{\beta}_{2n+1}$ (see
Eq. \eqref{betadip}) we can show after some algebra that 
the amplitude of a dip with frequency $\hat{\omega}_{2n+1,k}$ is given by
${\cal A}(\omega)\simeq \frac{F}{4\hat{\beta}_{2n+1}^8}
\left[\frac{\hat{\beta}_{2n+1}/2}{\tanh(\hat{\beta}_{2n+1}/2)}-1\right]^2
|\beta_c^4(\omega)-\hat{\beta}_{2n+1}^4|$. 
Hence, the minimum of the dip
\begin{multline} \label{Adipnk}
{\cal A}^{\rm dip}_{m,k} \simeq \frac{F L^4 \nu_t}{4{\cal E}_t}
\left(\frac{\hat{\beta}_{m}/2}{\tanh(\hat{\beta}_{m}/2)}-1\right)^2\times \\
\times \left[1+\frac{2Nm_t}{m_b}L(\gamma_{m,k})\right]
\frac{\hat{\omega}_{m,k}}{\hat{\beta}_{m}^8},
\end{multline} 
with $m=2n+1$, $n\geq 1$.

\subsubsection{Electromotive potential spectrum}

From Eqs. \eqref{y2solaverage} and \eqref{y2solaverageexact},
it is now straightforward to evaluate the electromotive potential $V_{emf}(t)$ 
given in Eq. \eqref{Vemf}. It reads 
\begin{equation} \label{Vemfsol}
\frac{V_{emf}(t)}{L B} = \frac{d \bar{y}(t)}{dt} 
= \omega_d {\cal A}(\omega_d) \sin 
(\omega_d t - \theta(\omega_d)-\pi),
\end{equation}
from which we eventually deduce the time r.m.s.
\begin{equation} \label{Vemfsolrms}
\sigma_{emf} \equiv 
\sqrt{\langle V^2_{emf}(t)\rangle} = \frac{BFL \omega_d}{\sqrt{2}} 
\frac{1}{|\beta_{c,d}^4|} \left|\frac{4T(\beta_{c,d})}{\beta_{c,d}}-1\right|.
\end{equation}  
From the formulas of the previous section we can evaluate 
the peak maxima and dip minima of the electromotive r.m.s. 
and we are able to compare them with those observed 
in experiments.

\section{Nonlinear system driven by two frequencies} \label{NLN2F}

It has been observed experimentally \cite{gaidarzhy_thesis} that, 
if in addition to being driven at the frequency of the fundamental mode 
the antenna is 
also driven at the frequency of the collective mode, the frequency 
peak of the fundamental mode experiences a slight shift. This frequency
shift is the signature of a modal coupling that occurs because of the
presence of nonlinearity and dissipation in the system \cite{Nayfeh04}. 
To explain the  
interaction of these two widely spaced modes, we investigate the effect 
of a two-frequency driving on the response of a weakly nonlinear and 
dissipative antenna. We supplement the equations
of motion of our continuum model, Eqs. \eqref{y2} and \eqref{eta2}, with 
nonlinear terms that take into account the possible 
material and geometrical nonlinearities of the structure and with damping 
terms proportional to the transverse velocity of the elements. 
These terms, small compared to the amplitude of vibration of 
the antenna, are treated within the Lagrangian approach described 
in \cite{Nayfeh04} to derive the frequency-amplitude relation of the model.

\subsection{Lagrangian approach}

In the previous section, we solved the linear continuum model
exactly and found its modes, 
$\varphi_{n,k} = (y_{n}(x,t),\eta_{n,k}(x,\xi,t))^T$, to be
related to two parameters: $\beta_n$,
root of $\cos \beta \cosh \beta =1$ and $\gamma_{n,k}$, solution to Eq. 
\eqref{beta4} or \eqref{secEq}. 
The frequency $\omega_{n,k}$ of $\varphi_{n,k}$ is determined by Eq. 
\eqref{Omega} and its modal shape derived from Eqs. \eqref{Huvsol} 
and \eqref{Yusol}. In particular, the fundamental mode is given by the
parameters $(\beta_1,\gamma_{1,1})$ and the collective mode we are 
interested in by the parameters $(\beta_1,\gamma_{1,2})$. For clarity
in the notations,
we rename these two modes $\varphi_1 = (y_1(x,t),\eta_1(x,\xi,t))^T$
and $\varphi_2 = (y_2(x,t),\eta_2(x,\xi,t))^T$, respectively and denote
their respective frequencies by $\omega_1$ and $\omega_2$ where
\begin{equation} \label{Omega1and2}
\omega_i = 
\sqrt{\frac{{\cal E}_t}{\mu_t}}\left(\frac{\gamma_{1,i}}{l}\right)^2, 
\ \ i \in \{1,2\}.
\end{equation}
It is observed experimentally that $\varphi_1$ and $\varphi_2$ 
are coupled in the sense
that the amplitude-frequency curve (resonance) of the fundamental mode
is modified when the higher mode is driven. This coupling is attributed 
to the presence of nonlinearities in the antenna. To explain this phenomenon,
we treat the nonlinearities and the damping affecting the system as a 
perturbation of the fundamental and collective modes, $\varphi_1$ and 
$\varphi_2$. These
perturbative terms are responsible for a modulation of the linear modes
that we evaluate by a multiple scale method. Hereafter, we closely follow 
the Lagrangian approach of Ref. \cite{Nayfeh04} because it offers 
a particularly suitable framework to derive the modulation equations. 
    
 When the driving amplitude 
(or power) is small enough, the solution $\varphi = (y(x,t),\eta(x,\xi,t))^T$
to the nonlinear equations of 
motion is, to a good approximation, given by a superposition of $\varphi_1$ 
and $\varphi_2$ with slowly modulated amplitudes. 
Following the multiple scale approach, we write it as 
$\varphi= A_1 \varphi_1+A_2\varphi_2$, that is
\begin{multline} \label{yetamultscal}
\begin{pmatrix} 
y(x,t) \\
\eta(x,\xi,t)
\end{pmatrix}
= \varepsilon\left\{ A_1(T_2)Y_1(u)\, e^{i\omega_1 T_0}\begin{pmatrix} 
1 \\
H_1(v)
\end{pmatrix} + \right. \\ \left. 
+ A_2(T_2)Y_1(u)\, e^{i\omega_2 T_0}\begin{pmatrix} 
1 \\
H_2(v)
\end{pmatrix} + c.c. \right\}
\end{multline}    
In this expression, $c.c.$ denotes the complex conjugate, 
$\varepsilon$ is a small bookkeeping parameter, $H_i(v)=\tilde{H}_{1,i}(v)$
where $\tilde{H}$ is defined in Eq. \eqref{Huvsol} and, 
according to the multiple scale method,
we have introduced two time scales, $T_0=t$ and $T_2=\varepsilon^2 t$.
The Lagrangian of our system is given by
\begin{multline} \label{Lagrangian}
{\cal L} = \int\limits_0^L dx 
\left[\frac{\mu_b}{2}\left(\frac{\partial y}{\partial t}\right)^2 -
\frac{{\cal E}_b}{2}\left(\frac{\partial^2 y}{\partial x^2}\right)^2  
\right] + \\ +
\frac{2N}{L} \int\limits_0^L dx \int\limits_0^l d\xi 
\left[\frac{\mu_t}{2}\left(\frac{\partial \eta}{\partial t}+
\frac{\partial y}{\partial t}\right)^2 - 
\frac{{\cal E}_t}{2}\left(\frac{\partial^2 \eta}{\partial \xi^2}\right)^2 
\right] + \\ 
+ ({\rm NLT}) + F_1 \cos(\Omega t) \int\limits_0^L y dx + 
F_2 \cos(\omega t) \int\limits_0^L y dx,
\end{multline} 
where $({\rm NLT})$ stands for ``NonLinear Terms''. 
To express the fact that the driving frequencies, $\Omega$ and $\omega$, 
are close to the linear frequencies of the fundamental and excited modes, 
we write them as
\begin{equation} \label{FreqDriv}
\Omega = \omega_1 + \varepsilon^2 \sigma_1\ \ ;\ \ \omega = \omega_2 + 
\varepsilon^2 \sigma_2.
\end{equation}   
To describe the nonlinear response of the cantilevers and the central beam, 
neglecting the effects of rotatory inertia and shear deformations, we
add the following nonlinear terms to the Lagrangian \cite{Nayfeh04}
\begin{multline} \label{NLterms}
\hskip-2ex {\rm NLT} \!=\! \int\limits_0^L \! dx \left\{\! \frac{\mu_b}{8}\! 
\left[\frac{\partial}{\partial t}\! 
\int\limits_0^x \!
\left(\frac{\partial y}{\partial x'}\right)^2 \!\!dx' \right]^2
\!\!\!-\! \frac{{\cal E}_b}{2} 
\left(\frac{\partial y}{\partial x}\,
\frac{\partial^2 y}{\partial x^2}\right)^2 \!\right\}+
\\
\frac{2N}{L}\!\int\limits_0^L \!\! dx\!\!
\int\limits_0^l \! d\xi \left\{ \!
\frac{\mu_t}{8} \!\! \left[\frac{\partial}{\partial t} 
\int\limits_0^\xi \!
\left(\frac{\partial \eta}{\partial \xi'}\right)^2 \!\!d\xi' \right]^2
\!\!\!\!-\! \frac{{\cal E}_t}{2} \left(\frac{\partial \eta}{\partial \xi}\,
\frac{\partial^2 \eta}{\partial \xi^2}\right)^2 \!\right\}.
\end{multline} 
Finally, we take into account the damping effects of the viscous 
forces acting on the antenna through the virtual
work
\begin{multline} \label{damping}
\delta W = - \int\limits_0^L \! dx \, 
C_y \frac{\partial y}{\partial t} \delta y 
\\
- \frac{2N}{L}\int\limits_0^L \! dx
\int\limits_0^l \! d\xi \, C_{\eta} \left(\frac{\partial y}{\partial t} + 
\frac{\partial \eta}{\partial t} \right) \delta \eta,
\end{multline}   
where $C_y$ and $C_{\eta}$ are the viscosities of the beam and the cantilevers,
respectively. 

To account for the fact that damping effects and driving forces are 
of the same order of magnitude as the nonlinear effects, we scale
the viscosities as $C_{y}=\varepsilon^2 c_{y}$ and 
$C_{\eta}=\varepsilon^2 c_{\eta}$ and the forces as $F_j = \varepsilon^3 f_j$,
$j=1,2$.
We now proceed as explained in Ref. \cite{Nayfeh04} to derive a time-averaged 
Lagrangian from Eqs. \eqref{yetamultscal}, \eqref{Lagrangian} and 
\eqref{NLterms}. We substitute \eqref{yetamultscal} into the Lagrangian 
\eqref{Lagrangian} and also into the virtual work \eqref{damping}, perform
the spatial integrations and keep the slowly varying terms only
- i.e. those that are either constant or function of $T_2$ only. This yields:
\begin{multline} \label{LagA}
\frac{\langle {\cal L} \rangle}{\varepsilon^4} = 
\sum_{j=1}^2 \left\{i M_j\omega_j \left(A_j \bar{A_j'}- A_j'\bar{A_j}\right)
+ C_{jj}|A_j|^4 + \right. \\ \left. + 
{\cal F}_j \left(\bar{A_j}e^{i\sigma_j T_2}+cc\right)\right\} 
+ 2 C_{12}|A_1|^2|A_2|^2 
\end{multline}
\begin{equation} \label{dWQ}
\frac{\langle \delta W \rangle}{\varepsilon^4} = \sum_{j=1}^2
Q_j \delta A_j + cc,\ \ \ {\rm with}\ \ \ Q_j = 2 i \omega_j \mu_j \bar{A_j},
\end{equation}
where
\begin{eqnarray} \label{Param}
M_j &=& m_b+2Nm_t L_{jj}\ \ ;\ \ {\cal F}_j = \frac{L}{2}f_j \Gamma_4 \nonumber \\
\mu_j &=& \frac{1}{2}( L c_y + 2Nl c_{\eta} \Lambda_{jj} ) \nonumber \\
C_{jj} &=& \frac{\Gamma_1 m_b \omega_j^2}{L^2} - \
\frac{3\Gamma_2{\cal E}_b}{L^5} +\frac{2N\Gamma_3}{l^2}
\left[m_t\omega_j^2I_{jj}-\frac{3{\cal E}_t}{l^3}K_{jjjj}\right] 
\nonumber \\
C_{12} &=& \frac{\Gamma_1 m_b (\omega_1^2+\omega_2^2)}{L^2}\!-\! 
\frac{6\Gamma_2{\cal E}_b}{L^5} \!+\! \frac{2N\Gamma_3}{l^2}
\Big[m_t(\omega_1^2+\omega_2^2)I_{12} \nonumber \\ 
&&  -\frac{{\cal E}_t}{l^3}
(K_{1122}+K_{2211}+4K_{1212})\Big]
\end{eqnarray}
and where
\begin{eqnarray} \label{integrals} 
\Gamma_1 &=& \int_0^1 \left(\int_0^u \left[Y_1'(\nu)\right]^2 d\nu \right)^2 
du ,\nonumber \\ 
\Gamma_2 &=&  \int_0^1 \left[Y_1'(u)Y_1''(u)\right]^2 du,\nonumber \\ 
\Gamma_3 &=&  \int_0^1 Y_1^4(u)\, du,\nonumber \\ 
\Gamma_4 &=&  \int_0^1 Y_1(u)\, du, \nonumber \\
L_{ij} & =& \int_0^1 h_i(v)h_j(v) \, dv,\nonumber \\  
\Lambda_{ij} & =& \int_0^1 h_i(v)(h_j(v)-1) \, dv,\nonumber \\ 
I_{ij} &=& \int_0^1 \left(\int_0^v h_i'(\nu) h_j'(\nu) \, d\nu \right)^2 dv, 
\nonumber \\  
K_{ijkl} &=&\int_0^1 h_i'(v) h_j'(v) h_k''(v) h_l''(v) \,dv.  
\end{eqnarray}
with $h_j(v) \equiv H_j(v)+1$. A numerical/analytical evaluation of the
above quantities for the antenna dimensions given in \ref{SecIIsub1} 
is provided in appendix \ref{AppInt}.

\subsection{Modulation equations}

Applying the extended Hamilton principle (see Ref. \cite{Nayfeh04}), we obtain
the equations of motion for the modulations $A_1(T_2)$ and $A_2(T_2)$ as
\begin{equation} \label{LagEq}
\frac{d}{dT_2} \left(\frac{\partial {\cal L}}{\partial \bar{A}_i'} \right) = 
\frac{\partial {\cal L}}{\partial \bar{A}_i} + \bar{Q}_i, \ i \in \{1,2\}.
\end{equation}
From Eqs. \eqref{LagA} and \eqref{dWQ}, we then derive the following pair of 
modulation equations
\begin{multline} \label{EqsA1A2}
2i\omega_i(M_i A_i'+\mu_iA_i) = -2A_i\left(C_{ii}|A_i|^2+ \right. \\
+ \left. C_{12}|A_j|^2\right)+{\cal F}_ie^{i\sigma_i T_2},
\end{multline}
where $(i,j) \in \{1,2\}$, $i \neq j$. 
Looking for solutions in the polar form
\begin{equation} \label{PolarAi}
A_i(T_2) = \frac{1}{2} a_i(T_2)\, e^{i(\sigma_i T_2-\theta_i(T_2))},  
\ i \in \{1,2\},
\end{equation}
and separating the real and imaginary components of Eqs. \eqref{EqsA1A2},
yields
\begin{eqnarray} \label{EqsA1A2realim}
\omega_i(\sigma_i-\theta_i')M_ia_i &=& 
\frac{a_i}{4}\left(C_{ii}\,a_i^2+C_{12}\,a_j^2\right)- {\cal F}_i 
\cos(\theta_i), \nonumber \\
\omega_i(M_ia_i'+\mu_i a_i) &=& {\cal F}_i \sin(\theta_i). 
\end{eqnarray}
Looking for steady state (periodic) solutions, we impose $a_j'=0$ and $\theta_j'=0$ and we finally obtain the frequency-amplitude relations as
\begin{eqnarray} \label{FreqAmp1}
\hskip-4ex \sigma_1\! &=& \!\frac{1}{4M_1\omega_1}\left[C_{11}\,a_1^2\!+\!C_{12}\,a_2^2\right]
\pm \sqrt{\frac{{\cal F}_1^2}{M_1^2\omega_1^2a_1^2}\!-\!\frac{\mu_1^2}{M_1^2}},
\\ \label{FreqAmp2}
\hskip-4ex \sigma_2 \!&=&\! \frac{1}{4M_2\omega_2}\left[C_{22}\,a_2^2\!+\!C_{12}\,a_1^2\right]
\pm \sqrt{\frac{{\cal F}_2^2}{M_2^2\omega_2^2a_2^2}\!-\!\frac{\mu_2^2}{M_2^2}}.
\end{eqnarray} 
together with
\begin{equation} \label{Anglestheta}
\tan \theta_i = \frac{4\mu_i\omega_i}{C_{ii}\,a_i^2+C_{12}\,a_j^2-
4M_i\omega_i\sigma_i},\ i \neq j.
\end{equation} 
In first approximation the steady state solution can be cast into the form
\begin{multline} \label{yetasol}
\begin{pmatrix} 
y(x,t) \\
\eta(x,\xi,t)
\end{pmatrix}
= Y_1(u) \left\{ a_1 \, \begin{pmatrix} 
1 \\
H_1(v)
\end{pmatrix} \cos (\Omega t - \theta_1) + \right. \\ 
\left. + a_2 \, \begin{pmatrix} 
1 \\
H_2(v)
\end{pmatrix} \cos (\omega t - \theta_2)  \right\}.
\end{multline}
The amplitudes $a_1$ and $a_2$ are assumed small enough for the 
perturbation expansion to hold (notice that the bookkeeping parameter
$\varepsilon$ has been absorbed in the amplitudes and that Eqs. 
\eqref{FreqAmp1}-\eqref{FreqAmp2} 
can be used as such provided the detunings $\sigma_j$ 
are redefined as $\sigma_j\equiv \varepsilon^2 \sigma_j$, the viscosities
$\mu_j$ as $\mu_j \equiv \varepsilon^2 \mu_j$ and the forces ${\cal F}_j$
as ${\cal F}_j \equiv \varepsilon^3 {\cal F}_j$).

\subsection{Discussion}

The frequency-amplitude relations \eqref{FreqAmp1}-\eqref{FreqAmp2}
allow us to evaluate the frequency shift of the fundamental peak 
induced by a driving of the
higher mode at the exact linear resonance frequency, $\omega_2$. This frequency
shift is determined as the difference between the maximum
of the resonance peak of the fundamental mode and $\omega_1$. Now, the 
amplitude$a_1$ becomes maximum if the square root in the r.h.s. of 
\eqref{FreqAmp1} vanishes, that is, 
\begin{equation} \label{a1max}
a_1^{\rm max} = \left| \frac{{\cal F}_1}{\mu_1 \omega_1}\right|.
\end{equation}
For a system whose higher mode is driven exactly at frequency $\omega=\omega_2$
we have of course $\sigma_2=0$ and then the amplitude of the second peak 
is solution to 
\begin{equation} \label{a2ofa1max}
\frac{1}{4M_2\omega_2}\left[C_{22}\,a_2^2+C_{12}\,
\frac{{\cal F}^2_1}{\mu^2_1 \omega^2_1}\right]
\pm \sqrt{\frac{{\cal F}_2^2}{M_2^2\omega_2^2a_2^2}-\frac{\mu_2^2}{M_2^2}} = 0,
\end{equation}
which is a cubic equation for $a_2^2$. Once the solution 
$a_2({\cal F}_1,{\cal F}_2)$
is known, we can reinstate it in \eqref{FreqAmp1} and we obtain the 
frequency shift, $\sigma_1 = \Omega -\omega_1$ as
\begin{equation} \label{Freqshifts1}
\sigma_1({\cal F}_1,{\cal F}_2) = 
\frac{1}{4M_1\omega_1}\left[C_{11}\,\frac{{\cal F}^2_1}
{\mu^2_1 \omega^2_1}+C_{12}\,a_2^2({\cal F}_1,{\cal F}_2)\right],
\end{equation} 
which provides the frequency shift as a function of the forces 
(or driving power), ${\cal F}_1$ and ${\cal F}_2$, of the fundamental 
and excited modes.

\section{Conclusion}

Here, we  have presented two analytical models  that yield closed-form
solutions  describing  the  dynamics  of the  coupled-beam  resonator,
dubbed  the antenna  structure.  This  structure is  a prototype  of a
class of  two-element mechanical structures that can  be envisioned in
specific applications that  involve coupled mechanical oscillators and
hierarchical structures.   The inherent modifications  associated with
the dynamics of coupled-element structures can be engineered to result
in  advantageous   frequency  and  amplitude   performance,  which  is
otherwise difficult  to obtain with simple  geometries. In particular,
the measurements  of a  similar nanomechanical fabricated  device have
been   reported   previously   \cite{gaidarzhy-prl,gaidarzhy-apl}   to
demonstrate some of the highest mechanical resonance frequencies up to
3 GHz, reported to date.

The continuum model allows for  a clear comparison of the modal shapes
and spectrum with full finite  element analysis of the structure.  The
findings   and  resulting   discussion  of   Sec.  \ref{SecComparison}
elucidate  the   behavior  of   the  coupled-element  system   in  the
fundamental and  first collective modes.   In particular, it  is shown
that the  enhanced effective amplitude of the  collective mode results
from  the collective excitation  of the  cantilever continuum  at high
frequencies, while  the supporting  central beam effectively  adds the
cantilever motion  by moving in  its fundamental mode shape  with zero
nodes,   thus  providing  maximum   transduction  of   the  cantilever
displacement to the measured magnetomotive voltage.

We have further investigated the driven damped model of the system as well
as the  nonlinear modal  coupling between widely  spaced modes  of the
structure using perturbation theory techniques.  The results elucidate
the  response of  experimentally  measured structures  that show  modal
interactions  even in  the limit  of linear  driving (to  be published
elsewhere). The analysis, however, is readily applied to a general set
of coupled element weakly  damped and weakly nonlinear resonators, and
it illustrates  the nontrivial modifications  in the dynamics  of such
systems that  can be carefully  engineered to suit  specific technical
needs.  As was  mentioned in the section \ref{Introduction},  
applications of RF
MEMS  and  NEMS  devices  are   numerous  in  the  areas  of  wireless
communications and frequency manipulation.

\begin{acknowledgements}
This work is supported by NSF (DMR-0449670).
\end{acknowledgements}

\appendix

\section{General solution to the discrete model} \label{Gensoldisc}
The solution to the system of equations \eqref{Yudiseq}-\eqref{Hjvdiseq} 
reads 
\begin{multline}
Y(u) = C^{(j)}_1 \cos(\mu u) + C^{(j)}_2 \sin(\mu u) + C^{(j)}_3 
\cosh(\mu u) + \\ + C^{(j)}_4 \sinh(\mu u) \ \ \text{for}\ \ 
u \in (u_{j-1},u_j),  \label{SolYudis}
\end{multline}
with $j \in \{1,\dots,N+1\}$, and 
\begin{multline}
\hskip-2ex H_i(v)= [A_1(\gamma) \cos(\gamma v) + A_2(\gamma) \sin(\gamma v) 
+ A_3(\gamma) \cosh(\gamma v) + \\ +  A_4(\gamma) \sinh(\gamma v)-1]Y(u_i),
 \label{SolHjvdis}
\end{multline}
with $i \in \{1,\dots,N\}$, $u_0=0$ and $u_{N+1}=1$. 
Given the boundary conditions \eqref{BCcant}, 
we can solve for $A_j(\gamma)$ in Eq. \eqref{SolHjvdis} 
and find the result given in Eq. \eqref{Aju}.
Now, let $C^{(j)}=(C^{(j)}_1,\dots,C^{(j)}_4)^T$. Using the boundary 
conditions \eqref{BCcant} for $Y(u)$ 
at $u_j$, we obtain \footnote{Because of Eqs. 
\eqref{SolHjvdis} and \eqref{Aju}, we have $Y'''(u_j^+)-Y'''(u_j^-) = -R
H_j'''(0)= 2\gamma^3 R A_2(\gamma)Y(u_j)= 4\mu^3 \alpha A_2(\gamma)Y(u_j)$, 
where $\alpha=w/W$.}
\begin{equation} \label{Cjp1vsCj}
C^{(j+1)} = (\One + 2\alpha A_2(\gamma) {\bf K_j}(\mu)) C^{(j)},\ 
j \in \{1,\dots,N\},
\end{equation} 
where the quantities involved are defined in section \ref{SecDiscVibMod}.
The boundary conditions Eq. \eqref{BCcant} at $u=0$ show that 
$C^{(1)}_3 = -C^{(1)}_1$ and $C^{(1)}_4 = -C^{(1)}_2$. The problem 
is then reduced to the two components $C^{(1)}_1$ and $C^{(1)}_2$. Rewriting 
them as the $2-$vector $c^{(1)}= (C^{(1)}_1,C^{(1)}_2)^T$, we have 
\begin{equation} \label{Extremu0}
C^{(1)} = {\bf L}\, c^{(1)},
\end{equation}
where ${\bf L}$ is defined in Eq. \eqref{MatKjLT}.
At the other end of the beam, $u=1$, the boundary conditions can also be cast
into a matrix form as
\begin{equation} \label{Extremu1}
{\bf T}(\mu)\, C^{(N+1)} = 0,  
\end{equation} 
where ${\bf T}(\mu)$ is again defined in Eq. \eqref{MatKjLT}. Putting
Eqs. \eqref{Cjp1vsCj}, \eqref{Extremu0} and \eqref{Extremu1} together, 
we finally see that 
\begin{equation} \label{Mc}
{\bf M}(\omega)\, c^{(1)} = 0.  
\end{equation}
where the $2 \times 2$ matrix ${\bf M}(\omega)$ is given in \eqref{MatM}. 
This system has a nonzero solution $c^{(1)}$ if and only 
if the determinant of ${\bf M}(\omega)$ is zero, hence the secular equation
\begin{equation} \label{SecEqDiscbis}
\det \left({\bf T}(\mu) \left[ \prod_{j=1}^N (\One + 2\alpha A_2(\gamma) 
{\bf K_j}(\mu)) \right] {\bf L} \right) = 0.
\end{equation} 
Once a solution $\omega_n$ of \eqref{SecEqDiscbis} has been found, 
the coefficients $C^{(j)}_i(\omega_n)$, $i=1,\dots,4$, 
of the corresponding mode, $Y_n(u)$, are automatically determined 
by Eqs. \eqref{Cjp1vsCj} and \eqref{Mc} as 
\begin{equation} \label{Cjomegan}
C^{(j+1)}(\omega_n) = \left[\prod_{k=1}^j 
(\One + 2\alpha A_2(\gamma_n) {\bf K_k}(\mu_n))\right] \,
C^{(1)}(\omega_n),
\end{equation} 
with $j \in \{1,\dots,N\}$ and 
\begin{equation} \label{C1omegan}
C^{(1)}(\omega_n) = {\bf L}\, c^{(1)}(\omega_n)\ \ \text{and}\ \  
c^{(1)}(\omega_n) = {\cal N} 
\begin{pmatrix} M_{12}(\omega_n) \\ -M_{11}(\omega_n) \end{pmatrix},
\end{equation} 
where $M_{ij}$ are the coefficients of the matrix ${\bf M}$. Ultimately, 
the normalization factor ${\cal N}$ is determined from 
$\int_0^1 \! Y_n(u)^2 \, du = 1$.

\section{Solution of the discrete model for $N=1$ (two cantilevers)} 
\label{App2tdeltavscont}
In this appendix, we find the solution of the discrete 
model for two cantilevers ($N=1$) located on both sides of the beam in its
middle and we briefly compare it to the solution of the continuum model
with the same number of cantilevers. Using the general result
provided in \eqref{SecEqDiscbis}, we can cast the secular equation into 
the form
\begin{equation} \label{SecEqDiscbisN1}
{\cal A}(\mu)\, {\cal S}(\gamma,\mu) = 0, 
\end{equation}
where 
\begin{equation} \label{AmuN1}
{\cal A}(\mu) =  \sin \frac{\mu}{2}\cosh \frac{\mu}{2} - 
\sinh \frac{\mu}{2}\cos \frac{\mu}{2}
\end{equation}
and
\begin{multline} \label{SmuN1}
{\cal S}(\gamma,\mu) = 2\alpha A_2(\gamma) 
(\cos \frac{\mu}{2}\cosh \frac{\mu}{2}-1)+ \\
+\sin \frac{\mu}{2}\cosh \frac{\mu}{2} + 
\sinh \frac{\mu}{2}\cos \frac{\mu}{2}.
\end{multline}
where $A_2(\gamma)$ is given in \eqref{Aju} and $\alpha=w/W$. 
The factorization of the
secular equation has a clear physical meaning. It is the result 
of the mirror symmetry of the problem with respect to the middle 
of the beam. Because of it,    
the modes of the discrete model are either symmetric ($Y_s$) or 
antisymmetric ($Y_a$) with respect to the latter.
In what follows, we use Eqs. \eqref{C1omegan} and \eqref{SecEqDiscbisN1}
to calculate the analytical form of the modal shape of the 
symmetric and antisymmetric modes for the central beam and the cantilevers. 
\subsection{Symmetric modes}
Symmetric modes satisfy $Y_s(u)=Y_s(1-u)$. Their secular equation 
is given by ($\mu > 0, \gamma > 0$), 
\begin{equation} \label{Secsym}
{\cal S}(\gamma,\mu) = 0,  
\end{equation} 
where ${\cal S}(\gamma,\mu)$ is given in \eqref{SmuN1} and, according to Eq. 
\eqref{nondim}, $\mu= \gamma L/l$. The corresponding frequency is 
obtained from
\begin{equation} \label{omegdis}
\omega=\sqrt{\frac{{\cal E}_b}{\mu_b}}\left(\frac{\mu}{L}\right)^2
=\sqrt{\frac{{\cal E}_t}{\mu_t}}\left(\frac{\gamma}{l}\right)^2.
\end{equation}
The expression for the modal shape of the central beam is,
\begin{multline} \label{Ysusol}
Y_s(u) = {\cal N}\Big\{\cos(\mu u) - \cosh(\mu u) + 
\left(\frac{\sin \frac{\mu}{2}+ 
\sinh \frac{\mu}{2}}{\cos \frac{\mu}{2}- 
\cosh \frac{\mu}{2}}\right) \times \\ 
\times (\sin(\mu u) - \sinh(\mu u))\Big\},\ u \in \left[0,\frac{1}{2}\right],
\end{multline}
with a normalization factor ${\cal N}$ given by
\begin{multline} \label{Normsym}
{\cal N} = \Big[1+\frac{6}{\mu}\frac{(\cos \frac{\mu}{2}
\cosh \frac{\mu}{2}-1)}{(\cos \frac{\mu}{2}-
\cosh \frac{\mu}{2})^2}\times \\
\times (\cos \frac{\mu}{2}\sinh \frac{\mu}{2}+
\cosh \frac{\mu}{2}\sin \frac{\mu}{2}) \Big]^{-1/2}.
\end{multline}
To evaluate the deflection of the cantilevers, 
we use Eq. \eqref{SolHjvdis} and find 
\begin{multline} \label{H1vsol}
H_1(v) = [A_1(\gamma) \cos(\gamma v) + A_2(\gamma)\sin(\gamma v) + \\ + 
A_3(\gamma)\cosh(\gamma v) + A_4(\gamma) \sinh(\gamma v) - 1]
Y_s(1/2),
\end{multline}
where the coefficients $A_i$ are given in Eq. \eqref{Aju} and where
the deflection of the middle of the beam is determined from \eqref{Ysusol} as
\begin{equation} \label{Yshalf}
Y_s(1/2) = 2{\cal N} \left(\frac{1- \cos \frac{\mu}{2}\cosh \frac{\mu}{2}}
{\cos \frac{\mu}{2}-\cosh \frac{\mu}{2}}\right).
\end{equation}
\subsection{Antisymmetric modes}
Antisymmetric modes satisfy  
$Y_a(1-u)=-Y_a(u)$, which leads to $Y_a(1/2)=0$. The deflection of the
cantilevers is then zero, $H_1(v)=0$. In this case, we directly see 
from equations \eqref{Yudiseq} and \eqref{BCcant}, 
that $Y_a(u)$ satisfies a simple clamped-clamped beam equation and that 
its third derivative has no discontinuity in $u=1/2$, given that $H_1(v)=0$. 
The shape 
of the antisymmetric modes is then similar to Eq. \eqref{Yusol}, which yields
\begin{multline} \label{Yausol}
Y_a(u) = \Big\{\cos(\mu u) - \cosh(\mu u) - \left(\frac{\cos \mu  
- \cosh \mu}{\sin \mu  - \sinh \mu }\right)\times \\ \times  
(\sin(\mu u) - \sinh(\mu u))\Big\},\ u \in [0,1].
\end{multline}
From Eq. \eqref{SecEqDiscbisN1}, we see that 
$\mu$ has to be a root of the secular equation 
\begin{equation} \label{Secantisym}
{\cal A}(\mu) = 0,
\end{equation}
where ${\cal A}(\mu)$ is given in \eqref{AmuN1}. It turns out that Eq. 
\eqref{Secantisym} is equivalent to the usual clamped-clamped secular 
equation, $\cos(\mu ) \cosh(\mu) = 1$, provided it is restricted to 
antisymmetric modes. Indeed, $\cos(\mu ) \cosh(\mu) = 1$
factorizes as ${\cal A}(\mu){\cal A}^{+}(\mu) = 0$ where 
${\cal A}^{+}(\mu)=\sin \mu/2 \cosh \mu/2 + 
\sinh \mu/2 \cos \mu/2$: ${\cal A}(\mu)=0$ provides
the frequencies of antisymmetric modes while ${\cal A}^+(\mu)=0$ gives
the frequencies of the symmetric modes of a simple clamped-clamped beam.
When cantilevers are affixed to the middle of the beam, symmetric
modes are affected by their motion and the secular equation
becomes ${\cal S}(\gamma,\mu) = 0$ (see Eq. \eqref{Secsym}) rather than
${\cal A}^+(\mu)=0$. Their modal shape changes from \eqref{Yusol} to 
\eqref{Ysusol}. For antisymmetric modes, however, the cantilevers do not move
so that, modal shape and frequency remain unaffected.    
 
The first roots of Eq. \eqref{Secantisym} 
are given by $\mu \simeq 7.8532$, $14.1371$, and  
$(2n+1/2)\pi$ as $n$ becomes large. The corresponding frequencies
are obtained from Eq. \eqref{omegdis}. Notice that, in this particular case,
the modal shape of an antisymmetric mode of the discrete model is exactly the 
same as the modal shape of the continuum model. Their frequency differs 
however.
Even though $\mu$ and $\beta$ (see Eq. \eqref{Eqbeta}) satisfy
the same secular equation (and are thus equal), 
the frequency of the continuum model is obtained 
by solving equation \eqref{beta4} whose solution, $\gamma_c$,  
is not proportional to $\beta$ while the solution of the discrete 
model, $\gamma_d$, satisfies $\gamma_d = \mu l/L$. The frequencies,
both obtained from Eq. \eqref{omegdis}, are thus different.
The exact similarity of the modes may seem surprising at first glance because,
if the cantilevers are at rest in the discrete model and thus 
do not participate
to the motion, in the continuum model, the force density of the cantilever has
been spread all over the beam and, consequently, the cantilever continuum 
moves with the beam. Nonetheless, as  
the force density it generates is everywhere proportional 
to the mode shape $Y(u)$, the frequency is detuned but the mode shape remains 
as is.   
\subsection{Comparison with the continuum model}
\begin{table*}
\caption{\label{tablegdisvsgcont} Values for the parameters 
$\gamma_d$ (discrete model) and $\gamma_c$ (continuum model) derived from
Eqs. \eqref{Secsym}-\eqref{Secantisym} and \eqref{Eqbeta}, respectively.
Frequencies are determined from Eq. \eqref{omegdis} for the antenna 
parameters given in the introduction. The symmetry class of the modes is
denoted by $s$ (symmetric) or $a$ (antisymmetric). }
\begin{ruledtabular}
\begin{tabular}{lcccccc}
Mode & Symmetry & $\gamma_d$ & $\gamma_c$ 
& $f_d$\footnote{The frequency is determined as $f=\omega/2\pi$} (MHz) & $f_c$\footnotemark[1] (MHz) & $\Delta f/f$ (in \%) \\
\hline
Fundamental & $s$ & 0.2149 & 0.2185 & 31.81 & 32.88 & -3.25 \\
First Excited & $a$ &  0.3670 & 0.3628 & 92.74 & 90.64 & 2.31\\
Second Excited & $s$ & 0.5035 & 0.5079 & 174.6   & 177.7 & -1.74 \\
Third Excited & $a$ & 0.6606 & 0.6530 & 300.5  & 293.4 & 2.33 \\
\end{tabular}
\end{ruledtabular}
\end{table*}
\begin{figure*}
     \centering
     \subfigure[]{
          \label{fig1}
          \includegraphics[width=.23\textwidth]{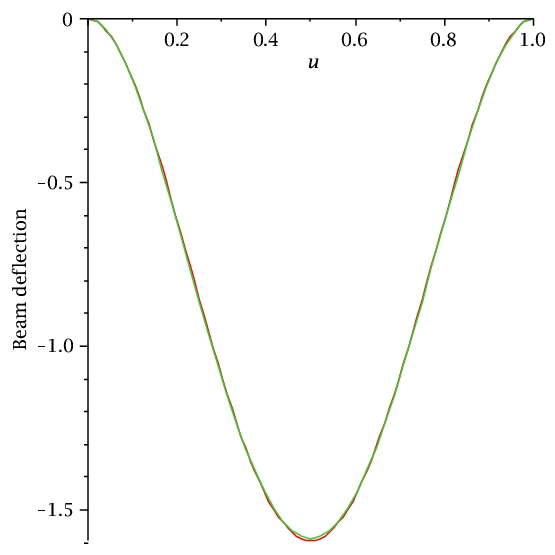}}
     \subfigure[]{
          \label{fig2}
          \includegraphics[width=.23\textwidth]{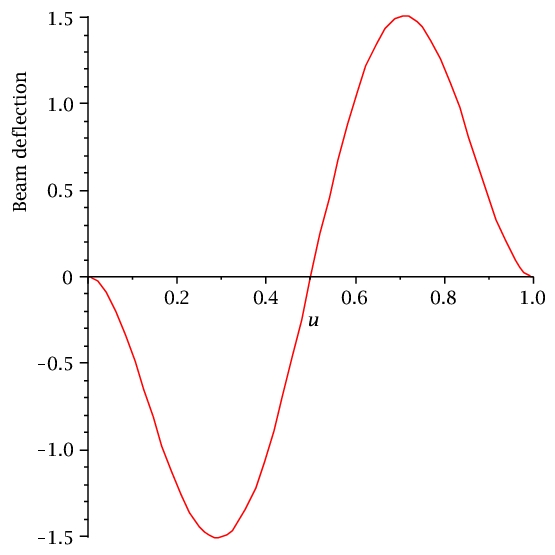}}
     \subfigure[]{
           \label{fig3}
           \includegraphics[width=.23\textwidth]{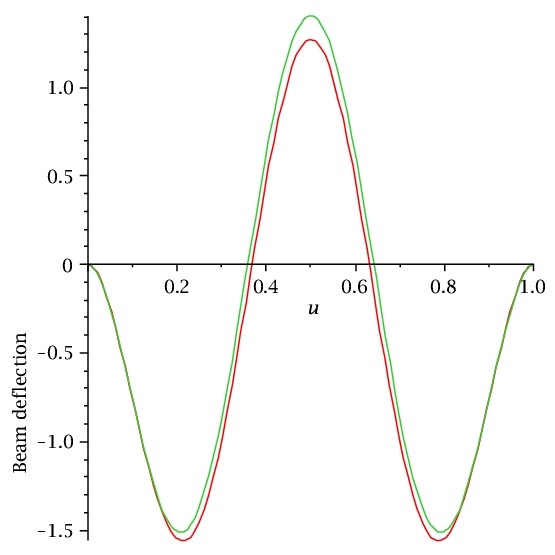}}
     \subfigure[]{
           \label{fig4}
          \includegraphics[width=.23\textwidth]{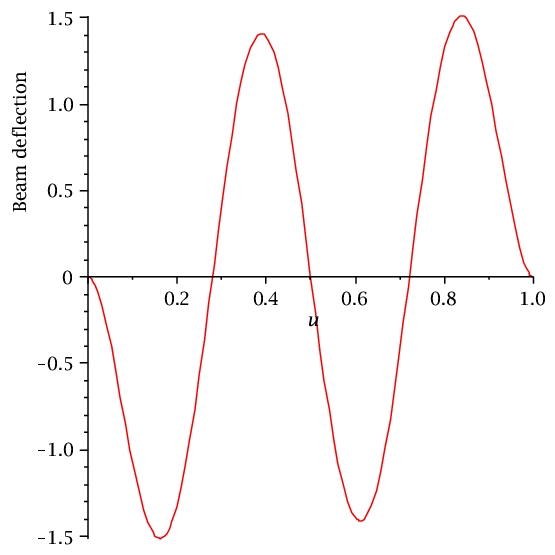}}\\
     \subfigure[]{
           \label{fig5}
           \includegraphics[width=.23\textwidth]{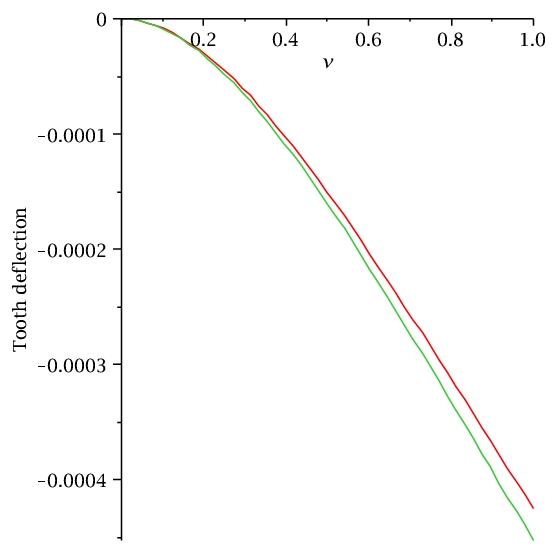}}
     \subfigure[]{
           \label{fig6}
          \includegraphics[width=.23\textwidth]{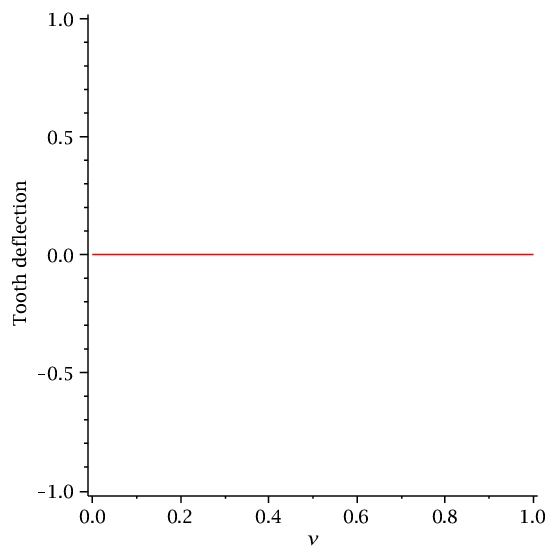}}
     \subfigure[]{
           \label{fig7}
           \includegraphics[width=.23\textwidth]{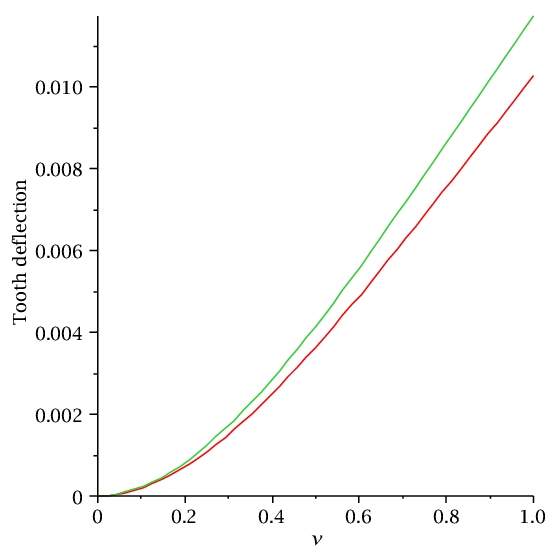}}
     \subfigure[]{
           \label{fig8}
          \includegraphics[width=.23\textwidth]{Disvscont_tooth_2ndand4th.jpg}}
     \caption{Fundamental to third excited mode shapes of the central beam 
((a),(b),(c),(d)) and of the 
cantilever ((e),(f),(g),(h)). The results of the discrete and continuum models are 
in red and green, respectively. For comparison, the deflection
of the cantilever continuum is shown at $u=1/2$ only. 
The antisymmetric discrete and continuum 
modes are exactly the same but their frequencies are different (see Table 
\ref{tablegdisvsgcont}).}
     \label{multifig}
\end{figure*}
According to the dimensions of the antenna, $L/l \simeq 21.4$ 
and $w/W \simeq 0.5$. Then $\mu \simeq 21.4 \gamma$ and $\alpha=0.5$. Plugging
these values into Eq. \eqref{Secsym}, we can solve for $\gamma_d$ and obtain
the values for the symmetric modes reported in Table \ref{tablegdisvsgcont},
namely $\gamma_{d,1}=0.2149$ and $\gamma_{d,3}=0.5035$. For the 
antisymmetric modes, we use $\mu \simeq 21.4 \gamma$ and the values quoted 
in the previous section and find $\gamma_{d,2}=0.3670$ and 
$\gamma_{d,4}=0.6606$.
Now, solving equation \eqref{Eqbeta} for $\beta=\beta_1$ to $\beta_4$, where
$\beta_n$ are the successive roots of $\cos(\beta ) \cosh(\beta) = 1$, we find
the corresponding values of $\gamma_c$ for the continuum model, reported 
in the third column of Table \ref{tablegdisvsgcont}. As we see,
these values are very close to those of the discrete model. 
Using Eq. \eqref{omegdis} we finally report the frequencies of the discrete 
(4th column) and continuum (5th column) modes. Their relative frequency 
difference is displayed in the 6th column. 
The modal shapes of the beam and the cantilevers from the fundamental 
mode up to the 
third excited mode are depicted in Fig. \ref{multifig}. For the cantilever 
continuum, we have displayed the deflection at $u=1/2$ only. As we can see, 
symmetric modes are very similar and antisymmetric modes are exactly the same.

Beside the first levels presented in table \ref{tablegdisvsgcont}, we have 
found numerically that the agreement between the spectra of the discrete 
and continuum models is very good in the sense that there is almost 
always one frequency of the continuum spectrum that closely matches 
the corresponding frequency of the discrete model. The shape of the central 
beam is very similar (same $n$) for matching frequencies. The frequencies
of the discrete model, however, are quite regularly spaced while in addition
to producing these regularly spaced frequencies, the spectrum 
of the continuum model also forms clusters of nearly degenerate frequencies
as explained in section \ref{SecFreqClus}. It is clear from the data 
we have obtained that, in the 
discrete model, the excitation level ($n$) of the central beam increases 
regularly with the frequency. In the continuum
model, however, this excitation level increases from $n=1$ to infinity
within each band, that is from $\omega_{1,k}$ up to $\omega_{\infty,k}$ 
determined by Eq. \eqref{Omega_infty}. Therefore, the continuum 
frequencies that match the discrete ones are located {\em away} from 
the band edge clusters of the continuum spectrum. The continuum frequencies
that accumulate close to the band edges have therefore no $(N=1)$ 
discrete equivalent.
In other words, collective modes such as those observed in the continuum
model appear only when the number of cantilevers is large enough for 
the continuum approximation to hold.

\section{Analytical and numerical results for Eqs. (\ref{integrals})} 
\label{AppInt}

Using Eqs. \eqref{Yusol} and \eqref{Huvsol},
all the integrals involved in the calculation of the effective parameters 
of the time average Lagrangian, see Eq. \eqref{integrals}, 
can normally be evaluated analytically. 
For $\Gamma_i$, $i \in \{1,\dots,4\}$, results are simple
enough to be displayed below. 
\begin{eqnarray}
\Gamma_1 & = & \frac{\beta_1 t}{2}(\beta_1 t+2) \simeq 6.1513, \nonumber \\
\Gamma_2 & =&  \frac{\beta_1^5t}{10}(5\beta_1t+11) \simeq 2846.4975, \nonumber \\
\Gamma_3 & =&  \frac{3}{4}\left(3-t^4-\frac{2t^3}{\beta_1}\right) 
\simeq 1.8519, \nonumber \\ 
\Gamma_4 &=& \frac{4t}{\beta_1} \simeq -0.8308 \nonumber 
\end{eqnarray}
where $t =\tan (\beta_1/2)$. Note that $\cos (\beta_1)=1/\cosh(\beta_1)$
and $\sin(\beta_1)=-\tanh(\beta_1)$.
Apart from $\Lambda_{ii}=L_{ii}-2A_2(\gamma_i)/\gamma_i$ and
$L_{ii}\equiv L(\gamma_{1,i})$, given in Eq. \eqref{Lnk}, 
the other integrals have been 
evaluated numerically for the parameters given in \ref{SecIIsub1}. 
We have found
\begin{eqnarray}
& &  L_{11} \simeq 1.00012 \ ;\ L_{22} \simeq 3.89887, \nonumber \\
& &  \Lambda_{11} \simeq 6.16\times 10^{-5} \ ;\ \Lambda_{22} \simeq 4.9687, 
\nonumber \\
& & I_{11} \simeq 1.64 \times 10^{-16} \ ;\ I_{22} \simeq 232.49, \nonumber \\
& & I_{12} \simeq 1.94 \times 10^{-7}\ ;\ K_{1212} \simeq 8.26 \times 10^{-7}, 
\nonumber \\
& &  K_{1111} \simeq 6.58 \times 10^{-16}\ ;\
K_{2222} \simeq 1.07 \times 10^{3}, \nonumber \\
& & K_{1122} \simeq 9.84 \times 10^{-7} \ ;\  
K_{2211} \simeq 6.98 \times 10^{-7}. \nonumber 
\end{eqnarray}
So that, finally,
\begin{eqnarray} \label{Paramnum}
&& M_1 = 1.74 \times 10^{-14}\ \ ;\ \  M_2 = 4.17 \times 10^{-14}, \nonumber \\
&& {\cal F}_i = -4.44 \times 10^{-6} f_i \ \ ;\ \ 
C_{12} = 1.65 \times 10^{17}, \nonumber \\
&& C_{11} = -5.07 \times 10^{13} \ \ ;\ \ C_{22} = 1.05 \times 10^{21}, 
\nonumber \\
&& \omega_1 = 1.55 \times 10^{8}\ \ ;\ \  \omega_2 = 1.85 \times 10^{10}. 
\nonumber 
\end{eqnarray}
where all the results are given in SI units.
From the quantities $\omega_j$, the frequencies of the 
fundamental and collective modes are $24.7$ MHz and $2.94$ GHz, 
respectively.

\end{document}